\documentclass[12pt]{article}

\usepackage{amsmath,amssymb,epsfig}

\begin{document}

\def\su2{$SU(2)$}
\def\sun{ $SU(N)$}
\def\C{{\mathbb C}}
\def\R{{\mathbb R}}
\def\tr{{\rm tr}}

\title
{\bf Properties of continuous Fourier extension \\
of the discrete cosine transform and \\
its multidimensional generalization}

\bigskip

\author{A.~Atoyan~~ and~~ J.~Patera \\
\small Centre de recherches math\'ematiques, Universit{\'e} de
Montr{\'e}al, \\
\small C.P. 6128, Succ.\ Centre-ville, Montr{\'e}al
(QC) H3C 3J7, Canada \\
}
\date{}

\maketitle

\begin{abstract}
A versatile method is described for the practical computation of 
the exact 
discrete  Fourier transforms (DFT), both the direct and the inverse ones, 
of a continuous function $g$ given by its values $g_{j}$ at the points of 
a uniform grid ${\sf F}_{N}$ generated by conjugacy classes
 of elements of finite adjoint order 
$N$ in the fundamental region $F$ of compact semisimple Lie 
groups. The present implementation of the method is for the groups 
$SU(2)$, when ${\sf F}$ is reduced to a one-dimensional segment, 
and for $SU(2)\times SU(2) \cdots \times SU(2)$ in multidimensional 
cases. This simplest case turns out to be a version of the discrete cosine 
transform (DCT). Implementations, abbreviated as DGT for 
{\it Discrete Group Transform}, based on simple Lie groups of higher 
ranks, are considered separately.

DCT is often considered to be simply a specific type of the standard DFT. 
Here we show that the DCT is very 
different from the standard DFT when the properties of 
the {\it continuous extensions} of the two inverse discrete transforms
from the discrete $t_j\in F_{N}$ to all points $t \in {\sf F}$
are studied. The following properties of the continuous extension of 
DCT (called CEDCT) are proven and exemplified.

\noindent
 Like the standard DFT, the DCT also returns the exact values of $\{g_j\}$ 
on the $N+1$ 
points of the grid. However, unlike the continuous extension of the standard 
DFT, 

\noindent
(a) the CEDCT function $f_N(t)$ closely approximates $g(t)$  
{\it between} the points of the grid as well.

\noindent
(b) For increasing $N$, the derivative of $f_N(t)$  
converges to the derivative of $g(t)$.

\noindent
(c) For CEDCT the principle of locality is valid.

\noindent
Finally we use the continuous extension of the 2-dimensional DCT,
$SU(2)\times SU(2)$,  to illustrate its potential for 
interpolation as well as for the data compression of 2D images.

\end{abstract}

\clearpage

\section{Introduction}

The decomposition of functions integrable on a finite segment into
Fourier series of trigonometric functions of one variable is a well
known method (e.g. \cite{Z,Tolst}) whose theoretical and practical
 aspects have been
thoroughly investigated during the last two centuries in connection
with its numerous applications in science and engineering. It is
natural to question whether any attempt to add something to it is
not in fact a reinvention of what has been found before.

Our general goal, which goes beyond this paper, is to elaborate a 
new decomposition method of functions of $n$ variables into
Fourier series using orbit functions of compact semisimple Lie groups 
\cite{MP1,MP4}, with the idea of (i) making it accessible to users who
are not specialists in Lie theory, (ii) underlining the versatility
of its practical implementations, and (iii) most importantly,
demonstrating the fertility of the underlying theme of this
approach. One can find complete answers to limited questions (like
the values of a finite number of Fourier coefficients) replacing
the Lie group by a suitably chosen set of its discrete elements.
 The choice of the discrete elements is clearly crucial. 

In the context of our goal, the $SU(2)$ 
case results in the simplest example of the new method, 
even though that is the case where the potential advantages of the 
method based on the symmetry groups could be most limited. 
Remarkably, however, in this low dimensional space the discrete 
Fourier transform on the SU(2) group results in one type of
famous discrete cosine transforms discovered in 1974 \cite{Rao1}, 
or more exactly the DCT-1, according to the currently accepted classification 
(see \cite{Wang,Strang}). Its comparison with the standard method is both 
misleading and revealing. It is misleading because it is considerably
similar (see \cite{DCT}) to the standard Discrete Fourier transform
 abbreviated typically as DFT (e.g. \cite{Brigham,OppenSch,Nuss}). 
Nevertheless it is revealing, since it can help
to understand better the underlying reasons why for many practical
applications the DCT is proven significantly more efficient than the DFT. 
In this paper we consider the concept of  {\it continuous extension} 
of the discrete transform, and show that the convergence properties of
the continuous extension of the inverse DCT, abbreviated here simply
as CEDCT, match very closely the properties of the canonical 
({\it continuous}) Fourier transform (CFT) of smooth functions. Meanwhile, the
continuous extension of the inverse DFT, abbreviated simply as
CEDFT, does not result in a reasonable function at all.
Note that for the sake of simplicity, 
if the ``inverse'' is not 
explicitly used, we will henceforth adopt DFT and DCT (or DGT
in a more general sense) abbreviations for both direct and 
inverse Fourier transforms.   

In Section 2  we present the basics of the Fourier analysis on 
the \su2 group, which also demonstrates the general formalism used 
for the Fourier transforms on Lie groups.    
We show that in practice the Fourier transform 
of a class function of \su2 to the orbit functions of this group is 
reduced to the decomposition of a discrete function $\{ g_{k} \mid
k = 0, 1, \ldots, N \}$ defined on the
$N$-interval grid of variable $t\in [0, T_0]$ onto the series 
of $(N+1)$ cosine functions (including $\cos 0 =1$) of the   
harmonic order $n=j/2 \leq N/2$. The basis for the DCT series is thus 
composed of the first $N$ {\it half-}harmonics of the $\cos(2\pi t/T_0)$ 
function, i.e. $\cos (2\pi n j t/T_0)=\cos (\pi j t/T_0)$,
meaning that the harmonic order of these functions may be formally both
integer for $j$ even,  {\it and } half-integer for odd $j$.
This approach is compared to the standard method of  DFT  
where the given $\{ g_{k}\}$ is decomposed into the  
trigonometric polynomials of $e^{(i \,2 \pi \, n t/T_0)}$
of the {\it full} harmonic order $n\leq N$. In Section 2 we also define
the continuous extension of the discrete transform on the continuum $[0,T_0]$.
We then compare CEDCT and CEDFT, and show that
although both DCT and DFT formally belong to the group of {\it exact} 
discrete transforms, surprisingly, only the CEDCT converges  
with increasing $N$ to the continuous function $g(t \in [0,T_0])$ 
which originates the grid function $\{ g_k \}$.

\vspace{2mm}

In Section 3, we prove some important properties of the CEDCT. 
These properties closely resemble those of the canonical  
 CFT polynomials where the coefficients are found 
by accurate integrations, such as the principle of  
{\it locality} of CEDCT. This feature can ensure, in particular, that  
numerical computation errors or uncertainties in one segment of the 
interval $[0, T_0]$ would not dramatically affect the results in 
a distant segment, which maybe very important for effective truncation
of the discrete transform sequence resulting in the loss of the 
{\it exactness} of the discrete transform. Another important property proven 
in Section 3 is that, similar to the CFT polynomials, the term-by-term 
derivative series
of the continuous extension of an $N$-interval DCT  converges to 
$g^\prime(t)$ with the increase of $N$. 
Note that this property holds for any smooth originating
function $g(t)$, in particular when $g_0 \neq g_N$ 
(which is not necessarily the case for other types of discrete Fourier 
transforms).  

In Section 4, we extend the formalism of one-dimensional DGT on \su2,
or the DCT, for decomposition of multidimensional functions, and bring
examples of approximation of some 2-dimensional discrete functions/images 
by a continuous extension of 2-dimensional CEDCTs.

\section{Basics of Fourier analysis on SU(2)}

The Lie group \su2 can be realized as a set of all 2$\times$2 
unitary complex-valued matrices $A$, with $\det A = 1$.
A complex valued {\it class function\/} $f$ on \su2 is any map of
\su2  onto the complex number space ${\C}$ which is invariant
under  conjugation,  i.e. $f : \su2 \longrightarrow \C$, and
$f(B^{-1} A \, B)=f(A)$ for all $A,B\in\su2 $.   Since the defining
2-dimensional representation of \su2 is faithful, we can use it in
order to describe the discrete elements of \su2 of interest to us.

Any unitary matrix can be  diagonalized by a unitary
transformation. Therefore every element of \su2 is conjugate to at
least one diagonal matrix in the defining 2-dimensional
representation. All the elements which can be simultaneously
diagonalized form a maximal torus {\sf T} of \su2. Since all maximal
tori are \su2-conjugate, we can write:
\begin{equation}\label{torus}
{\sf T} = \left\{ x(\theta) = \left( 
        \begin{array}{cc}
          e^{2\pi i \theta} & 0 \\
         0  & e^{- 2\pi i \theta} \\
         \end{array} \right) \; \mid \; 0\leq \theta \leq 1 
    \right\} \; .
\end{equation}
Furthermore, using 
$B= \left( \begin{smallmatrix}0&1\\-1&0\\
\end{smallmatrix} \right)   \in\su2 $, 
we have $Bx(\theta)B^{-1}=x(-\theta)$. Therefore  every element of
\su2 is conjugate to just one element in the subset $F\subset T$,
where
\begin{equation}\label{F-space}
{\sf F} = \left\{ x(\theta) = \left( 
        \begin{array}{cc}
          e^{2\pi i \theta} & 0 \\
         0  & e^{- 2\pi i \theta} \\
         \end{array} \right) \; \mid \; 0\leq\theta \leq \frac{1}{2} 
    \right\} \; .
\end{equation}

Trace functions, otherwise called {\it characters}, play an important role 
in \su2. For any element $x(\theta)$ of \eqref{torus}, we
have
$\tr\,x(\theta)=2\,\cos(2\pi \theta)$.  In general,
$\tr\,x(\theta)$ is a class function because for all 
$B\in\su2\Rightarrow\tr\,\{B^{-1}x\,B\} = \tr\,x$. Let
$R=R(\su2)$ denote a complex algebra generated by the character 
functions of \su2. It is well known then that $R$ has a linear basis 
consisting of the characters of all  finite dimensional irreducible 
representations of \su2.       

With each irreducible  representation of a semisimple Lie group,
in particular \su2, one  associates a set of weights (weight
system) of the representation \cite{MP4,MP1}, which is a union of  
orbits of weights under the action of the Weyl group, $W$. 
In physics the
\su2-weights are known as projections of the angular momenta which 
have integer and half-integer eigenvalues. In mathematics one usually
prefers  to deal with the doubles of the angular momenta in order to avoid
non-integers.

The Weyl group of \su2 is very simple. It is of order 2, consisting of
2 elements generated by action of the reflection operator $\hat{r}$. 
It acts on any element $m\in\mathbb{R}$ of the 1-dimensional space
$\mathbb{R}$ of the `projections of angular momenta' in the
straightforward way: $\hat{r}(m)=-m$.
The finite dimensional irreducible representations of \su2 and
of its Lie algebra $su(2)$ are
well known, but we would point out the following. The `angular momentum
states', the basis vectors of representation spaces, are eigenvectors
of the `diagonal' generator of $su(2)$. Unlike the common
normalization of that generator in physics, we normalize it so
that its eigenvalues are twice the usual projections of angular
momenta. The set of the eigenvalues $\Omega(l)$ designates the
weight system of the representation $l$. The weight system of an
irreducible representation 
consists of $l+1$ weights,
$$
\Omega(l) = \{ m\omega\mid m\in\{-l,-l+2, \dots,
l-2,l\}\,\}\,, 
$$
where $l$ is the highest weight (`twice the angular momentum') of
the representation\footnote{For sets of weights in case of Lie
groups  different from \su2 see \cite{MP1}}; 
$l$ is used to specify the representation. Note that
all the elements of $\Omega(l)$ have the same parity.

A $W$-orbit of a weight $m$ thus consists of one or two elements: 
$$
Wm= \left\{ \begin{array}{cl}
            \{m, -m\} & {\rm for} \; m\neq0 \\
                 \{0\} &  {\rm for} \; m= 0 \\
            \end{array} \right. \; .
$$

The character $\chi(\theta)$ is a function of conjugacy classes of
the elements of \su2. Every class is represented by one value of
$\theta$ within $0\leq\theta\leq\frac{1}{2}$. The values of the
character of an irreducible representation $l$ can be written as the
sum of values of the orbit functions  $\Phi_m(\theta)$,
\begin{equation}\label{character}
\chi_l(\theta) = \sum_m^l \Phi_m(\theta) = 
\Phi_l(\theta)+\Phi_{l-2}(\theta)+\cdots+
\left\{ \begin{array}{ll} 
       \Phi_1(\theta),   & l \ \rm{odd}\\
       \Phi_0(\theta), & l\ \rm{even}
         \end{array}
\right. \,. 
\end{equation}
\begin{equation}\label{orbit}
\Phi_{m}(\theta) = 
\left\{ 
\begin{array}{ll}
   e^{2\pi im\theta} + e^{-2\pi im\theta} &
                               {\rm for} \; m>0\\
   1 & {\rm for} \; m=0 
\end{array}
\right.\,, \qquad \rm{and\quad} 0\leq\theta\leq\frac{1}{2}\,.
\end{equation}
Only for the 1- and 2-dimensional representations, $l=\, 0$ and 1
respectively, the character consists of a single orbit function.
Note that
$\Phi_{m}(\theta)$ is symmetric (antisymmetric) with respect to
the midpoint of its range of 
$\theta$  for $m$ even (odd).

The decomposition of irreducible characters \eqref{character}
into the sum  of orbit functions is given by a triangular matrix.
Hence it is invertible. Therefore the orbit functions
$\{\Phi_m(\theta),\ m=0,1,2,\dots\}$ also form a basis in the
space of class functions $f(\theta)$ of \su2. Using
\eqref{orbit}, it implies that 
\begin{equation}\label{expansion}
f(\theta) = \sum_{m=0}^{\infty} a_{m}\,\Phi_{m}(\theta) \;
= a_0+2\sum_{m=1}^\infty a_{m}\,\cos(2\pi m\theta),
\qquad 0\leq\theta\leq\frac{1}{2}\,.
\end{equation}
There are two properties of the expansion \eqref{expansion} which
we want to underline:  \\
(i) \ It can be reduced to the familiar case of the standard
Fourier  decomposition of $f(\theta)$ in the interval $\theta \in
[-1/2,1/2]$, if one makes an even extension
$f(\theta) = f(-\theta)$ for $t\in [-1/2,0]$. \\
(ii) Although $f(\theta)$ is being expanded into a series of
functions which are periodic within the range $0\leq\theta\leq1$,
the actual range of $\theta$ in \eqref{expansion} makes periodic
only the cosines with even values of $m$, i.e $m=2k$. Their
arguments vary over the range $\{0,2k\pi\}$, i.e. over an integer
multiple of $2\pi$.

\subsection{Discrete Fourier transform on \su2} 

The discrete Fourier transform differs from \eqref{expansion} by
the fact that the independent variable $\theta$ takes only finite number of
 rational values within its range of variation.

Fixing a rational value of $\theta$, one fixes an element of finite
order (EFO) belonging to the \su2 torus {\sf T}. Every conjugacy
class of EFO in \su2 is represented by an element of {\sf T} with
$0\leq\theta\leq\frac{1}{2}$. In \su2 one can be explicit, see
\cite{Kac} or \cite{MP2}\ \S4\ for all other compact simple Lie
groups.

Let ${\sf T}_N$ denote the set of all elements of {\sf T} whose
adjoint order divides $N$, where $N$ is a positive integer.
The adjoint order is the order of the element represented by matrices
of irreducible representations of \su2 of odd dimensions (i.e. $l$
even). There are exactly $(N+1)$  \su2-conjugacy classes of such
elements. Taking the unique diagonal matrix as representative of
each conjugacy class, in representations of dimensions $2$
and $3$, we have the following set of matrices
$$
{\sf T}_{N}^{(2)} = \left\{ x_{N,k} = \left( 
        \begin{array}{cc}
          e^{\frac{2\pi i k }{2 N} } & 0 \\
         0  & e^{- \frac{ 2\pi i k}{2N} } \\
         \end{array} \right) \; \mid \; k=0,1,\ldots,N 
    \right\} \; ,
$$

$$
{\sf T}_{N}^{(3)} =  \left\{  x_{N,k} =
       \left(\begin{array}{ccc} 
        e^{\frac{2\pi ik}{N} } & 0 & 0 \\
           0 & 1 & 0 \\
        0  & 0 & e^{- \frac{ 2\pi ik}{N}} \\
         \end{array} \right)
\mid \; k=0,1,\ldots,N
    \right\} \; .
$$

The trace functions of each of these matrices, of size $(l+1)\times
(l+1)$, represent the characters $\chi_l(\theta)$ of the representation 
$l$, which can be used for decomposition of the class functions 
on \su2 (and generally on compact simple Lie groups).

A more suitable basis for such a decomposition
consists of orbit functions \cite{MP4,MP1}. It  makes possible
({\it practical}) decomposition of class functions on groups of high rank,
e.g. such as $E_8$ \cite{MMP1,MMP2,GP}. In the case of \su2, the
orbit functions are also much closer to the familiar  set of
exponentials $\exp (2\pi im\theta)$ used in the standard Fourier
analysis. 

Note that the elements $x_{N,k}$ are equidistant 
over the fundamental region ${\sf F} = \{\theta\in[0,1/2]\}$ of the
Weyl group $W$. The number of elements of the $W$-conjugacy class
$m=k/N$ is denoted by $C_{N,k}$ and is given by
\begin{equation}\label{classsize}
C_{N,k} = \left\{  \begin{array}{ll}
1 & {\rm if} \; k=0, N \\
2 & {\rm otherwise} \\
\end{array}
\right. \; .
\end{equation}

The following definition of
a sesquilinear form $\langle f,g\rangle_{N}$ in  the space $R$ of
class functions $f$ and $g$ on \su2  is a crucial step for our method:
\begin{equation}\label{sform}
\langle f,g\rangle_{N} = 
\sum_{k=0}^{N} C_{N,k} f(x_{N,k}) \,\overline{g(x_{N,k})}\; ,
\end{equation}
where the overline stands for complex conjugation.
It is known \cite{MP1,MP2}, and it can be verified by direct
computation, that the set of orbit functions 
$\{\Phi_k\mid k = 0,\ldots,N\}$ is orthogonal on the discrete equidistant  
$N$-interval grid with respect to this form. More precisely, we have:
$$ 
\frac{\langle\Phi_k,\Phi_m\rangle_N}{\langle\Phi_m,\Phi_m
      \rangle_N} = \delta_{km}\;,\qquad  
\text {for}\qquad 0\leq k,m\leq N. 
$$

For further convenience and for comparison of the results  with
the conventional Fourier series, let us instead of orbit
functions $\Phi_{m}(\theta)$ defined by \eqref{orbit} consider the functions
$\psi_m(\theta) =  \cos(2\pi m\theta)$ for all $m\geq 0$. Then one
can easily verify that for $0 \leq k, m \leq N$ 
\begin{equation}\label{orthogonality}
\langle \psi_k , \psi_m \rangle_{N} = \sum_{j=0}^{N} C_{N,j} \, 
\cos \frac{\pi j k}{N} 
\cos\frac{\pi j m }{N} =\frac{2\,N}{C_{N,k}}\delta_{k m} \; ,  
\end{equation}
where $C_{N, k}$ is given by \eqref{classsize}. 
The method proposed below for decomposition of the class functions
into series  of orbit functions is based on the discrete
orthogonality relation \eqref{orthogonality}. 

 Let $f(\theta)\in R$ be a class function that can be 
decomposed into a {\it finite} series of orbit functions:
\begin{equation} \label{discrete}
f(\theta) =\sum_{m=0}^N a_m \, \psi_{m}(\theta)\,, \qquad
( 0\leq \theta \leq\frac{1}{2})\,. 
\end{equation}
This can be compared with the general case of infinite
series \eqref{expansion} and with the discrete Fourier transform
(2) in \cite{B}.

  Our goal now is to find the expansion coefficients $\{ a_m \} $. 
In order to use the orthogonality property \eqref{orthogonality},
we form a system of
$(N+1)$ linear equations for 
$\{a_m\}$, restricting $\theta$ in \eqref{discrete} to the
discrete set of its values
$\{\theta_{k} = k\Delta\theta\,
\mid k= 0,\ldots,N\}$, with $\Delta\theta =\frac{1}{2N}$ :
\begin{equation}\label{expanded}
f_k\equiv f(\theta_{k}) 
= \sum_{m=0}^{N} a_m\psi_{m}(\theta_k)
= \sum_{m=0}^{N} a_m\cos \frac{\pi m k}{N}\,.
\end{equation}
After multiplication of \eqref{expanded} by $C_{N,k} \,\psi_{j k}$
and summing over $k$, we arrive on the right hand side at 
$\sum_{m=0}^{N}a_m\langle\psi_{m},\psi_{j}\rangle_{N}$. 
Then, given \eqref{orthogonality}, we find 
\begin{equation}\label{ajSU2}
a_{j}= \sum_{k=0}^{N} D_{N}^{j k } f_{k} \; \; 
{\rm for}\;j = 0,\ldots,N\;,
\end{equation}
where
\begin{equation}\label{Djk}
  D_{N}^{j k} = \frac{C_{N,j} \, C_{N,k} }{2 \, N} \, \cos 
\frac{\pi k j }{N} \;  .
\end{equation} 
Here $D_{N}^{j k}$ are the elements of $(N+1)\!\times\!(N+1)$
matrix $D_{N}$ of the DGT on SU(2). Note that it is easily reduced to the
transform matrix of the discrete cosine transform of the type DCT-1  
after renormalization by a factor $\sqrt{2 C_{N,j} \, C_{N,k}/ N }$.
The matrix $D_N$ is
independent of the values $\{f_k\}$ of the class function which is
being decomposed. It is therefore possible to compute $D_N$ in
advance,  for any predefined values of the positive integer $N$,
and use it repeatedly whenever it is needed.

Examples of the transform matrices $D_N$ for the lowest values
of $N$ are the following:
\begin{eqnarray}\label{D3}
D_1 =    \left( \begin{smallmatrix}
 \frac{1}{2} & \frac{1}{2} \\ \frac{1}{2} & - \frac{1}{2} 
          \end{smallmatrix}  \right),\;  
D_2 = \frac{1}{2} {\left(\begin{smallmatrix}
 \frac{1}{2}  & 1 & \frac{1}{2}\\ 1& 0 & -1 \\
        \frac{1}{2} & -1 &  \frac{1}{2} 
          \end{smallmatrix}\right)},\;  
D_3 = \frac{1}{3}{ 
      \left(\begin{smallmatrix}
           \frac{1}{2}& 1 & 1 & \frac{1}{2}\\
           1 & 1 & -1 & -1 \\
           1 & -1 & -1 & 1 \\
          \frac{1}{2} & -1 & 1 & -\frac{1}{2}
          \end{smallmatrix}\right)},\, 
         \\ 
D_4 = \frac{1}{4}{\left( \begin{smallmatrix}
            \frac{1}{2} & 1 & 1 & 1 & \frac{1}{2} \\
           1 & \sqrt{2} & 0 & -\sqrt{2} & -1  \\
         1 & 0 & -2 & 0 & 1 \\
         1 & -\sqrt{2} & 0 & \sqrt{2} & -1 \\
          \frac{1}{2} & -1 & 1 & -1 & \frac{1}{2} 
         \end{smallmatrix}\right)\; .} \hspace{3.3cm}  
\nonumber 
\end{eqnarray}

Introducing single-column matrices   $A_N = \{a_j\}$ and  
$F_N = \{f_{N}(t_k)\}$, and a square matrix
$\Psi_N =\{\cos(\pi jk/N)\mid  j,k=0,\ldots,N\}$, 
equations \eqref{expanded} and \eqref{ajSU2} are written in 
the matrix form, 
\begin{equation}\label{matrixDGT}
F_{N}=\Psi_{N}A_{N}\;\quad {\rm and}\quad\;A_{N}=D_{N}F_{N}\,,
\quad\text{where}\quad D_N=\Psi_{N}^{-1}\,.
\end{equation}
The matrix $D_N$, being the inverse of $\Psi_N$, formally 
solves the problem of discrete Fourier transform on \su2.

\subsection{Continuous extension of the discrete Fourier transforms} 

Let a continuous function $g(t)$  be the origin for the 
discrete function $\{g_k =g(t_k)\}$ defined 
 at the $(N+1)$ points $t_k = k T_0/N$,  $k=0, 1, \ldots N$,
 of the interval $[0,T_0]$. The DCT of $\{ g_k\}$ with the use of 
the transform matrix \eqref{Djk} results in the discrete function  
$\{a_j\}$ in the frequency space. This is an exact (`lossless') 
discrete transform, since it allows unambiguous recovery of all $N+1$ values
of $\{g_k\}$ by applying the inverse DCT in the form of \eqref{expanded}.
We recall that the standard discrete Fourier transform, i.e. DFT, 
has the same property. 

It seems then natural to ask if it is possible to recover
the originating function $g(t)$ by a Fourier series not only at the grid 
points $\{ t_k\}$, but also  on the entire continuous 
segment $[0,T_0]$. In order to answer this question, let us consider   
the continuous extension of the discrete transform between the grid points,
which can be formulated as follows.

\smallskip

\noindent
{\bf Proposition:} {\it Let $g(t)$ be a complex valued function of
$t\in[0, T_0]$, taking values $g_k =g(t_k)$ on an equidistant
point grid $\{t_k= kT_0/N\,\mid k= 0, 1, \ldots,N\}$. The function 
\begin{equation}\label{CEDGT} 
f_{N}(t\in [0,T_0])= \sum_{j=0}^{N}a_j\cos(\pi jt/T_0) 
\end{equation}
with the discrete transform coefficients 
\begin{equation}\label{CEDGTinv}
a_j=\sum_{k=0}^{N}\frac{C_{N,j}C_{N,k}}{2N} g_k
\cos\frac{\pi j k}{ N}\, ,
\end{equation}
represents a continuous Fourier extension of the  
inverse DCT of $\{ g_k\}$, which is exact in the sense that 
$f(t_k) = g_k$ at all $N+1$ points of the grid.}

The proof of this proposition is obvious
if one recalls that \eqref{CEDGT} and \eqref{CEDGTinv} are  reduced,
respectively, to \eqref{expanded} and \eqref{ajSU2} after
substitution $t=2\theta T_0$ with   $\theta\in[0,\frac12]$.    

\vspace{2mm}

Similar to DCT, one can also continuously extend to all points of the segment
$[0,T_0]$ any other types of discrete transforms, in particular the
DFT, resulting in CEDFT. 

Below we will address two important questions. First, how well 
the CEDCT approximates any `reasonably behaved'
(e.g. continuous) function $g(t)$ on the interval $[0, T_0]$ 
outside the points of the grid.  The
second question is how it compares with the CEDFT. At last, we will also 
briefly address the question of the possible use of CEDCT in practical 
applications, in particular, for purposes of smooth representation 
of compressed images.

\vspace{2mm}

In order to provide a partial answer to the first question, we
consider two examples. 

\noindent{\bf Example~1.} 
Let us take a Gaussian function
\begin{equation}\label{Gauss}
g(t) = \exp[-\frac{1}{2}(t/\sigma)^2],\qquad t\in [0,1],
\end{equation}
with the dispersion $\sigma=1/3$, and choose $N=3$.
Thus we have chosen a rather coarse grid relative to the
dispersion: the width of its intervals 
$1/N$ is equal to the dispersion. The coefficients \eqref{ajSU2}
are readily calculated  using $D_3$ from \eqref{D3}:
\begin{equation}\label{A3}
A_3 = \begin{pmatrix} a_0\\a_1\\a_2\\a_3\end{pmatrix} 
=\frac13\begin{pmatrix} 
  0.5 +e^{-0.5} +e^{-2} +0.5 \, e^{-4.5}\\
  1 +e^{-0.5} - e^{-2} -  e^{-4.5} \\
  1 - e^{-0.5} - e^{-2} + e^{-4.5} \\
  0.5 +e^{-0.5} +e^{-2} +0.5 \, e^{-4.5} \\
          \end{pmatrix} 
=\begin{pmatrix} 0.415807 \\
           0.486695 \\
          0.089748 \\
         0.007750 \\
 \end{pmatrix} 
\end{equation}
The corresponding CEDCT function \eqref{CEDGT} reads:
\begin{equation}\label{f3}
f_3(t)=a_0+a_1\cos(\pi t)+a_2\cos(2\pi t)+a_3\cos(3\pi t),
\end{equation}
where the coefficients are given by \eqref{A3}. 

Now we are in a position to compare \eqref{Gauss} with
\eqref{f3}. At the grid points $t_k=k/3$ the functions coincide,
$f_3(t_k) = g(t_k)$ (which is easily verified). One may then 
expect large deviations of
$f_3(t) $ from $g(t)$ around intermediate points of the grid, i.e.
$t=1/6,\, 1/2, $ and 5/6\,:
$$
\begin{pmatrix}f_3(1/6)\\f_3(1/2)\\f_3(5/6)\\\end{pmatrix}= 
\begin{pmatrix}0.882171\\0.326059\\0.039191\end{pmatrix}
\quad\longleftrightarrow \quad
\begin{pmatrix}g(5/6)\\g(1/2)\\g(1/6)\\\end{pmatrix}= 
\begin{pmatrix}0.882497\\0.324652\\0.043937\end{pmatrix}. 
$$
A good agreement between $f_3(t)$ and $g(t)$ is apparent 
even for $t = 5/6$ where $g(t)$ becomes very small.

\medskip

\noindent
{\bf Example~2.\/}
DCT approximation for a more complicated  function is found in
Fig.~1, where  we show $f_{N}(t)$, $N=10$ and $14$, for the
function $g(t)$  composed of two Gaussians,
\begin{equation}\label{figure1}
g(t)= A_1 e^{-\frac{1}{2}\left(\frac{t-t_1}{\sigma_{1}}\right)^2 } +
A_2 e^{-\frac{1}{2}\left(\frac{t-t_2}{\sigma_{2}}\right)^2 } \; ,
\end{equation}  
with amplitudes $A_1=2$, $A_2=1.5$, narrow dispersions
$\sigma_1=\sigma_2=0.05$, and centered at $t_1=0.42$,
$t_2=0.56$.

For comparison, in Fig.~1 we show by dashed lines  
the approximations to $g(t)$ provided by the
trigonometric CFT polynomials 
where the transform coefficients are calculated by  exact integrations. 
Recall that for a real function $g\rightarrow\R$ 
the CFT polynomials of the harmonic order $K$ are given 
by the series\footnote
{
In general, for a complex function $g(t)$ the  coefficients 
$c_{-j} \neq \bar{c}_{j}$.
} 
(see e.g. \cite{Tolst}):   
\begin{equation}\label{exact}
P_{K}(t)= \sum_{j=-K}^{K} c_{j} \, e^{i 2\pi j t/T_0}
= c_0 + 2 \, {\rm Re} \sum_{j=1}^{K} c_{j} \, e^{i 2\pi j t/T_0} \; ,
\end{equation}
where   
\begin{equation}\label{trigoninv}
c_j = \frac{1}{T_0}\int_{0}^{T_0} g(t) \, e^{-i 2\pi j t/T_0} \, {\rm d} t
\; .
\end{equation}
In order to compare approximations to $g(t)$ that can be provided
by series \eqref{CEDGT} and \eqref{exact} with  the same order
for the highest harmonics, for the series $P_K(t)$ in Fig.1 we put
$K=N/2$.

Figure~1 illustrates that the DCT  is really an exact
discrete Fourier transform, i.e. that $f_N(t_j)=g(t_j)$ for all
$0\leq j\leq N+1$. Remarkably, its continuous extension
$f_N(t\in [0,T_0])$ approximates $g(t)$ practically as well as
the accurate  
CFT trigonometric Fourier series $P_K(t)$ of the same harmonic order,
i.e. $K=N/2$.  In the case of Gaussian-type functions, CEDGT series
approximates the original function $g(t)$ reasonably well even in
the case of narrow structures with dispersions as small as
$\sigma \approx T_0/1.5 N$.

\subsection{Comparing with standard DFT}

First, let us recall some properties of the standard  exploitation
of the discrete Fourier transform. Further details can be found
in many books (e.g. see \cite{Brigham, OppenSch, Nuss}).

The standard DFT is formally derived from an approximate
calculation of the integral coefficients $c_j$ for the
trigonometric Fourier series, using a simple rule of rectangles for
integration of $c_j$ in \eqref{trigoninv} when the function $g$ is 
given on the $N$-interval equidistant grid, 
$g(t)\rightarrow\{g_{k}\mid k= 0,\ldots N\}$. This leads to DFT
coefficients   
\begin{equation}\label{stDFT} 
u_j=\frac1N\sum_{k=0}^{N-1}g_ke^{-i\frac{2\pi}{N}kj}\,,   
\end{equation} 
where $1/N$ is the length of the sampling interval $\Delta t$
(assuming  for simplicity $T_0=1$). 

A crucial feature of this definition of DFT 
consists in the fact that the system of equations \eqref{stDFT} for the
first $N$ coefficients 
$\{u_j\mid j=0,\ldots,N-1\}$ can be inverted with respect to 
$\{g_k\mid k=0,\ldots,N-1\}$. Such a possibility is based on the
observation that the matrices with matrix elements 
$$
M_{jk} = \frac1N e^{-i\frac{2\pi}{N}kj}
\qquad\text{and}\qquad
(M^{-1})_{jk} = e^{i\frac{2\pi}{N}kj}
$$  
are inverse to each other. Thus one gets the {\it inverse\/}
DFT in the following form
\begin{equation}\label{stDFTinv}
g_k = \sum_{j = 0}^{N-1} u_{j} e^{i \frac{2 \pi}{N} k j} \; .
\end{equation}
Thus, the discrete sets $\{ u_{j} \}$ and $\{ g_k \}$
represent a pair of exact ({\it lossless}) direct and inverse transforms
(e.g. see \cite{Nuss}) in the form of Fourier series,
which is generally treated as the exact {\it solution} to the
problem of Fourier transform of discrete functions given on 
the equidistant grid. 

We argue below, however, that there are significant reasons
to suggest that the real (and not only formal) exact solution to the
problem of discrete Fourier transform of a grid function is provided by the  
DCT transform pair given by \eqref{CEDGTinv}
and \eqref{expanded}. It retains all the `good' properties of the DFT:

(i)~~it is easy and fast to compute ;

(ii) it is a lossless discrete transform, with the exact inverse DCT
at all $(N+1)$ points of the grid (even if $g_0 \neq g_N$, unlike for DFT); 

However, in addition to these properties, the continuous extension
of DCT, $f_N(t)$ in \eqref{CEDGT}, converges to the originating 
continuous function $g(t)$ with increasing $N$,
as illustrated in the previous section, and as it is proved in the next
Section. Only the continuous extension of the DCT, and not the DFT, 
reveals properties characteristic to the canonical continuous Fourier 
transform series.

It is worth noting here that the very good convergence properties 
seem to be a common feature for the discrete Lie group transforms, 
as we also demonstrate on the example of SU(3) group in the 
accompanying paper. The basic mathematics of DGT has been
formulated \cite{MP2} for any dimension $n<\infty$. In fact there
are as many different variants of the method one could use, as
there are different semisimple compact Lie groups of rank $n$, and
then within each variant the choice of the points of the grid is
also far from unique, except for the lowest cases like \su2.
It then provides an opportunity to make a choice of appropriate DGT in 
situations where the choice of symmetry is dictated by the 
experimental data.

The absence of the convergence property for the continuous
extension of the DFT given by
\eqref{stDFTinv} is not easy to anticipate because CEDFT looks like 
a Fourier polynomial, or a cut-off of an `ordinary' Fourier expansion: 
\begin{equation}\label{CEDFT}
h_{N}(t)=\sum_{j=0}^{N-1}u_{j}e^{2i\pi\,jt}\,.
\end{equation}

Similar to CEDCT, it satisfies the equality
$h_{N}(t_k)=g_k$ on the grid points $t_k=k/N$ for\footnote
{
Note  that the last $k=N$ grid knot can also be included in DFT
provided that $g(0) = g(T_0)$.
}
all $k\leq N-1\,$. The fact that their  continuous extensions CEDGT
and CEDFT behave quite differently is illustrated in 
Example~3. It has rarely been emphasized that 
$h_{N}(t)$ {\it does not} approximate the initial
function  $g(t)$ between the grid points, and it does not converge
at all to any continuous function (except for a trivial case of 
$g(t)= const$) with increasing $N$. It is worth citing in this
regard \cite{OppenSch}, p.~87: "the DFT is a Fourier 
representation of finite length sequence which is itself a sequence
rather  than a continuous function". 
\smallskip

\noindent
{\bf Example~3.}
Let $h_N(t)$ be the CEDFT of N-interval grid function arising from 
a sampling of the continuous function  
\begin{equation}\label{figure2}
g(t)=A e^{-\left( \frac{t-t_0}{a} \right)^6} \; ,
\end{equation}
with  $A=2$ and parameter $a = 0.15$. Here we have chosen a large 
value, $6$, for the power in the exponent in order to illustrate a
case with gradients significantly larger than in the case of
Gaussian functions.

Solid lines in  Fig.~2 correspond to continuous DFT extensions
$h_{N}(t)$, and the dashed lines show the DCT extension
$f_{N}(t)$.  Although $h_{N}(t)$  passes through all $g_k$ at
$t=t_k$ (shown as full dots),  similar to $f_N(t)$, its behaviour
in between shows profound oscillations due to the
presence  in \eqref{CEDFT} 
of high-frequency Fourier components, $\omega_j= 2\pi j/T_0$
(if $T_0\neq 1$), with values of $\frac{N}{2} \lesssim j\leq N$
comparable to $N$. These oscillations in CEDFT 
do not decrease with increasing $N$, but they quickly disappear in CEDCT. 

This behaviour is explained by the fact that at any large $N$ 
the $j$-th order harmonic, $\exp(-2i\pi jt)$, strongly varies
and changes its sign
in a narrow interval $\Delta t = 1/N$ for high orders 
$j\gtrsim N/2$. Therefore the rectangular
integration rule in \eqref{trigoninv} cannot provide any reasonable
similarity between the canonical CFT coefficients $c_j$ in
\eqref{trigoninv} and their standard DFT `approximations' $u_j$ in
\eqref{stDFT}.  Effectively, the `fine tuning' between the coefficients
for high-order harmonics intrinsic to the continuous Fourier transform
is lost. 

\subsection{Comments on Fast Fourier Transform}

There is a number of ways to make a significantly more accurate
approximation of the high-frequency coefficients $c_j$  for a
function given on the discrete grid. Despite this fact, 
the standard version of DFT defined by \eqref{stDFT} and
\eqref{stDFTinv} for the direct and inverse  discrete transforms
has been widely used since the  pioneering paper by Cooley and
Tukey \cite{CoolTuk}, where the first algorithms for fast calculation of 
this DFT were developed. Different algorithms for such {\it fast Fourier
transform} (FFT) computations allow an increase in the
speed of practical calculations of DFTs  by one or two orders of
magnitude. Thus, a direct `head-on' algorithm for  calculations of
$\{ u_j \mid j=0, \ldots , N-1\} $ would require about $N^2$ 
multiplications and additions. Meanwhile, for special values of $N$,
FFT algorithms can significantly reduce the required number of
elementary operations, e.g. down to $\sim N \log_{2} N$ in  
case of   $N=2^n$. 

The discussion of various FFT  algorithms, extensively developed later
on by many authors  (e.g.
\cite{Gentleman,Singl,RadBren,Winograd}), is outside
 the scope of this article. Here we note only that the FFT
methods are fundamentally exploiting  the property of the
standard DFT coefficients that in their complex-value  representation
\eqref{stDFT} they can be reduced to a  power-law series of a single
complex element $W_{N}= \exp(-i 2\pi/N)$
as $u_{j} =N^{-1} \sum g_k W_{N}^{jk}$. This would not be possible 
if more accurate integration methods to deal with the
high-order harmonics were used. It should be noted, however, that 
the DCT does allow for the application of FFT methods since it can be formally 
reduced to $2N$-point DFT, and a number of efficient FFT-based algorithms
have been developed for DCT (e.g. \cite{NarPet, TsengMil, Mak, VetNus}). 
Moreover, a number of efficient algorithms competing with FFT and 
specific to DCT have been developed (see \cite{DCT} for details).

A simple modification could significantly improve 
 behaviours of continuous extensions of discrete Fourier
transform series based on the use of standard DFT coefficients 
\eqref{stDFT}, but not without penalty. Namely, recall that 
for a sufficiently smooth
function $g(t)$ in \eqref{trigoninv}, the rectangular integration
rule provides good accuracy for approximating $c_j$ for 
low-order  harmonics, $j\leq N/2$. Since FFT algorithms allow fast 
calculation of $\{ u_j\}$, and as far as $g(t)$ can be generally
well approximated with the CFT trigonometric polynomials
\eqref{exact} with  $K\leq N/2$ (as shown below), one can only use
the first half of the  standard DFT coefficients $u_j$ for
construction of a continuous extension of the inverse discrete
transform sequence.
It will then be similar to the series \eqref{exact} truncated to
harmonics of order $j\leq N/2$, i.e. it represents a series 
similar in structure to the CEDFT sequence \eqref{stDFTinv},
but where the high order harmonics are eliminated:    
\begin{equation}\label{approximate}
 s_{K}(t)= u_0 + 2 {\rm Re} \sum_{j=1}^{K} u_{j} \, e^{i 2\pi j t/T} \; ,
\end{equation}
where the coefficients $u_j$ are defined by \eqref{stDFT}.
Note the difference in the multiplication factor 
2 in this series at $j \geq 1$ as 
compared with the standard DFT extension $h_{N}(t)$ of \eqref{stDFTinv}.  

In Fig.~3 we show that the function $s_{K}(t)$ with $K=N/2$ (dashed
curves) can indeed approximate $g(t)$ practically as well as the DCT
extension $f_{N}(t)$.
But in this case the penalty is a loss of the `exactness' property for such
modified DFT sequence. That is, $s_{K}(t) \neq g_{k}$ at every $t=t_k$ for 
$k\leq N+1$. Therefore \eqref{approximate} cannot represent 
an exact solution to the problem of discrete Fourier transform.
Meanwhile, the series $f_{N}(t)$ both satisfies that condition,
 and rapidly  converges to $g(t)$ with increasing $N$. 

For comparison, we also show in Fig.~3 (dot-dashed line)  
the function $s_K(t)$ calculated for $K=6<N/2=8$. In this
case the approximation errors $[s_K(t)-g(t)]$ are larger than 
the ones when $K=8$, because the order of high frequency harmonics 
becomes important for the approximation of features with a dispersion
$\sigma \leq T/2K$.

\section{ Localization and differentiability of CEDCT}

In this section we prove the properties {\it localization} and 
{\it differentiability} of the CEDCT which
are analogous to the properties of the canonical CFT 
polynomials \eqref{exact} (e.g. see \cite{Tolst}).  

Derive first a useful formula for  an N-interval CEDGT
\eqref{CEDGT}
of a grid function $\{g_k\}$. Using \eqref{CEDGTinv}, and assuming for 
simplicity $T_0 = 1$, i.\,e. $t\in [0,1]$, \eqref{CEDGT} is reduced
to 
\begin{equation}\label{CEDGTun}
f_{N}(t) =\frac{1}{2 N} \sum_{j=0}^{N} \sum_{k=0}^{N} C_{N,k} C_{N,j}
          \cos(\pi j t) \, \cos (\pi j t_k) \, g_{k} \; ,
\end{equation}
where $t_k \equiv k/N$. Using 
$2\cos\alpha\cos\beta =\cos(\alpha +\beta)
+ \cos(\alpha - \beta)$, the terms containing index $j$ can be
re-written as a sum of two geometric series:
$$
\sum_{j=0}^{N} C_{N,j} \cos (\pi j t) \, \cos (\pi j t_k) =
(-1)^{k}\cos (\pi N t) -1 
+ {\rm Re} \left( \sum_{j=0}^{N-1} e^{i \pi (t+t_k) j} +  
\sum_{j=0}^{N-1}e^{i \pi (t-t_k) j} \right)  
$$
Summing up the series, one gets a compact expression for CEDGT:
\begin{equation}\label{DGTsum}
f_{N}(t) = \sum_{k=0}^{N} \frac{(-1)^{k} C_{N,k}}{2N} \, 
\frac{\sin \pi N t \, \sin \pi t}{\cos \pi t_k - \cos \pi t}
\: g_k\, 
\equiv \sum_{k=0}^{N} A_{N,k}(t) \, g_k , 
\qquad  {\rm  for} \quad t\in [0,1] \; . 
\end{equation}
The values of $f_N(t) $ for any 
$t\rightarrow t_n =n/N$ are found by applying the Lipshitz
rule for the ratio of infinitesimals of a smooth function. For 
\eqref{DGTsum} it results    
in $f_{N}(t_n) = g_n$ for all $n = 0, \ldots , N$, as expected.

Using \eqref{DGTsum}, below we prove that the {\it
localization principle\/}, known (see \cite{Tolst}) for the CFT 
series \eqref{exact} also holds for the CEDCT. The localization
lemma can be formulated as follows: 
\medskip

\noindent
{\bf Localization Lemma.} {\it Let the set of $N$-interval
grid functions $\{ g_k\}$, with various $N$, be originated from 
a smooth function $g(t)$ with a bounded derivative $g^\prime (t)$ 
on the interval $[0, 1]$. 
Then at any given $t\in [0,1]$ and for any small $\Delta >0 $ and
$\epsilon > 0$ there exists $ N_{\epsilon,\Delta}$ such that for all  
$ N > N_{\epsilon,\Delta}$ the behaviour of the continuous extension
of the discrete group transform  $f_{N}(t) $  
is defined within the accuracy $\epsilon$ only by the values
of $g(t)$ in the $\Delta$-neighborhood of $t$, i.e.:

for any $ \epsilon, \Delta >0$~~~$\exists N_{\epsilon,\Delta}$ such that
for all $ N > N_{\epsilon, \Delta} \Longrightarrow $
$$
\mid f_{N}(t) - \sum_{\{k^\prime\} } A_{N,k^\prime}(t) g_{k^\prime}
\mid < \epsilon
\quad   ,\, with \; all \; k^\prime  \; within \;  \;
t-\Delta < \frac{k^\prime}{N } < t+\Delta \; . $$ }

\vspace{2mm}
\noindent
{\bf Proof:} From \eqref{CEDGTinv}, 
using \eqref{orthogonality}, it follows that 
the DGT coefficients of any constant function $g_1(t) = const$ are 
equal to $a_{k}^{(1)} = \delta_{k 0} \times const$, i.e. that all
coefficients 
except for $a_{0}^{(1)} = const$ are equal to 0. Thus the convergence 
properties of the CEDGT of any function $g(t)$ are the same as the properties
of the function $g(t) + const$. So taking into account that 
the function is bounded, it is sufficient to prove the Lemma, while generally 
assuming that $g(t)$ is positive on the interval [0,1]. 

Let us split the sum in \eqref{DGTsum} into 3 parts corresponding to 
\begin{eqnarray}\label{Spm} 
 S_{-} & = & \sum_{k= 0}^{K_t - K_{\Delta}-1} A_{N,k} \, g_k \; ,
\nonumber \\
 S_{+} & = & \sum_{k= K_t + K_{\Delta}+1}^{N}  A_{N,k} \, g_k \; ,\\
 S_{\Delta} & = & \sum_{k = K_t - K_{\Delta}}^{K_t + K_{\Delta}}
 A_{N,k} \, g_k \; . \nonumber
\end{eqnarray}
where
$K_t = [N t]$ and $K_{\Delta} = [ N \Delta]$ are the integer parts of
the respective products.
If any of the points $t \pm \Delta$
is outside the interval [0,1], then only 2 sub-series are left. 
In all cases, only $S_{\Delta}$ of these sub-sums is defined by $g(t)$ in
the close neighborhood of $t$. 
The localization Lemma for \eqref{DGTsum} then implies that the {\it
non-local} sums, $S_{+}$ and $S_{-}$, are reduced with  increasing N
to absolute values below any small $\epsilon$. 

Indeed, on the basis of
\eqref{DGTsum}, for any fixed $\Delta$ each  of those non-local sums
represents a series of bound-value elements with alternating signs, 
which can be then combined into pairs of consequtive elements 
of order $O(1/N^2)$ each. Considering for example the sum $S_{-}$, 
and using the Taylor series decomposition for $g_{k+1} \equiv g(t_{k+1})
= g_k + g^{\prime}(t_k)/N +o(1/N)$ each pair of elements 
$ A_{N,k} g_k + A_{N,k+1} g_{k+1} $ starting with even $k\geq 2$ 
can be reduced to
$$
\frac{\sin \pi N t \, \sin \pi t}{N^2}
\left[ \frac{g^{\prime}(t_k)}{\cos \pi t_{k} - \cos \pi t}
- \frac{g_k \sin \pi t_k}{(\cos \pi t_{k} - \cos \pi t)^2} \, + \, o(1/N^2) 
\right].
$$ 
The expression in the square brackets is bounded
with some absolute value independent of $N$ as far as $g^{\prime} $ is  
bounded and $\mid t_{k}-t \mid > \Delta$, therefore 
$A_{N,k} g_k + A_{N,k+1} g_{k+1} \sim O(1/N^2) $. 
The number of such pairs in $S_{-}$ or $S_{+}$ is increasing  $\propto N$.
Therefore taking also into account that in the  sums $S_{-}$ and $S_{+}$
each of the limited number of elements left out from such pairing process (e.g.
in case of $C_{N,k} = 1$ for $k=0,\, N $) is only of order $O(1/N)$,  
we conclude that both series $S_{-}$
or $S_{+}$ tend to zero with the increase of $N$. This proves the localization
Lemma.  
\hfill$\blacksquare$

\vspace{4mm}
A very important property of the CEDGT series $f_N(t)$ is the possibility 
of its term-by-term differentiation such that the resulting series
converges  with increasing $N$ to the derivative $g^{\prime}(t)$. 
Note that while being well-known for the CFT series \eqref{exact}, 
this property is not trivial for finite (N-)element discrete 
Fourier transforms. Recall that although in trigonometric
polynomials in the form of \eqref{CEDGT} or \eqref{approximate}
each individual term, being $\propto 1/N$, decreases to 0 
at large N,  their derivatives over $t$ corresponding to a 
high-order harmonics, say $j> N/2$, become of order
$j/N \sim 1 $, and therefore might not necessarily vanish with 
increasing $N$. Thus, a `fine tuning' of the entire 
discrete-transform based series is needed in order to
provide convergence of the series produced by its term-by-term 
differentiation.

\vspace{2mm}
\noindent
{\bf Theorem:} {\it Let $g(t)$ be a smooth function with bounded second
derivative on the interval $t \in [0,T_0]$, which originates the
N-interval grid function $\{g_k \mid k=0,1,\ldots , N\}$. 
Then the function
$f_{N}^{\prime}(t)$ produced by the term-by-term differentiation of 
the continuous extension of the discrete Fourier transform on SU(2),  
$f_{N}(t)$, converges with increasing $N$ to $g^{\prime}(t)$ 
at any $t\in (0,T_0)$. }
    
\vspace{2mm}

\noindent
{\bf Proof}.  As earlier, we put $T_0=1$ for simplicity of the formulae
below.
Because the series $f_{N}(t)$ given by \eqref{CEDGTun} 
contains a finite number of elements, it is obvious that its
derivative $f_{N}^{\prime}$ can be summed up to the series 
produced by the term-by-term differentiation of  \eqref{DGTsum}. 

Consider first the derivative $f_{N}^{\prime}(t)$ at any 
{\it rational } $t_0$ in the open interval (0,1). In that case we
can choose $N$ such that $t_0=m/N$, and it then makes sense to choose   
all further subdivisions of the interval [0,1] such that 
$t_0$ will be always kept as a knot of the grid, i.e. 
$t_0 = m_1/N_1$ for all $N_1 > N$, i.e. choosing $N_1 = a N$ with some
integer $a>1$. Using the Lipshitz 
rule, the derivative of \eqref{DGTsum} at $t_0=m/N$ is reduced to
\begin{equation}\label{fprime}
f_{N}^{\prime}(t_0)  =  \frac{\pi}{2} 
\sin \pi t_0 \sum_{k=0}^{N \; (k\neq m)}
(-1)^{k-m}\, C_{N,k} \, U_{k}(t_0) \; + \;
 \frac{\pi \cos \pi t_0}{2 \sin \pi t_0} \, g_m \; ,
\end{equation}
\begin{equation}\label{Ukt0}
U_{k}(t_0) = \frac{g_k}{\cos \pi t_k - \cos \pi t_0}  \; ,
\qquad {\rm where }\;  t_k = k/N \; .
\end{equation}

Let us choose some small $\Delta$, such that both $(t_0\pm \Delta) \in
(0,1)$,
and then split the sum in \eqref{fprime} into 3 sub-series 
$S_{+}^{\prime}, \, S_{-}^{\prime}$ and
$S_{\Delta}^{\prime}$ as in \eqref{Spm}, where the number 
$K_t=m$ for $t=t_0=m/N$.
It is convenient for further analysis to choose $K_\Delta$
as the maximum integer which satisfies the condition $K_\Delta/N \leq
\Delta$ and is of the same parity as $m$. Then the indices of both the 
last element in the series $S_{-}^{\prime}$ and the first element in 
$S_{+}^{\prime}$,  $m-K_\Delta -1$ and $m+K_\Delta +1$ 
respectively,  are odd.

 Recall that for any smooth function $U(t)$ with a bounded second 
derivative on the equidistant grid we have  
\begin{equation}\label{Uk2}
2 U_k = U_{k-1} + U_{k+1} + U_{k}^{\prime \prime} \, N^{-2} 
+ o(N^{-2}) \; ,
\end{equation}
which follows from the familiar Taylor series decomposition 
$g(t+x) = g(t) +g^{\prime}(t)\, x +  0.5 \, g^{\prime \prime} (t)  
\, x^2 + o(x^2)$ in the
$x=\pm 1/N$  vicinity of $t$. Applying \eqref{Uk2} to 
$U(t)=g(t)/[ \cos \pi t - \cos \pi t_{0}]$ at all $t=t_k$ 
in \eqref{fprime} and \eqref{Ukt0} with $k$ odd (i.e.     
$k=2 j -1 \mid 1\leq j
\leq (m-K_\Delta)/2 \,$), and given \eqref{classsize}, 
the series $S_{-}^{\prime}$  is reduced to 
\begin{equation}\label{Sminprime}
S_{-}^{\prime} = (-1)^{m} \frac{\pi \sin \pi t_0}{2} 
\left(- U_{m-K_{\Delta}} + N^{-2} 
\sum_{j=1}^{(m-K_\Delta)/2} U_{2j -1}^{\prime \prime}\right)
 + o(N^{-1}).
\end{equation}
The first term on the right-hand side is exactly one half of the first
term
 in the localized sum $S_\Delta$, but with a negative 
sign. The second term is
of order $O(N^{-1})$. A more precise estimate of this term can be derived
if we note that for large $N$ the sum 
\begin{equation}\label{Uestimate}
\frac{2}{N} \sum_{j=1}^{(m-K_\Delta)/2} U_{2j-1} \rightarrow 
\int_{0}^{t_0-\Delta}
U^{\prime \prime}(t) {\rm d} t = U^{\prime}(t_0-\Delta)-
U^{\prime}(0) \; , 
\end{equation}
is bounded for any given $\Delta$, as far as 
$g^{\prime\prime}(t)$ is bounded. Note that the
estimate of \eqref{Uestimate} implies only that  
$U^{\prime\prime}(t)$ is integrable. This suggests that in
principle the conditions of the validity of the differentiation
Theorem can be relaxed, requiring that $g^{\prime\prime}(t)$ be
an integrable function on $[0,1]$, but not necessarily bounded.   

Applying the same approach to $S_{+}^{\prime}$, we find 
\begin{equation}\label{Spmprime}
S_{-}^{\prime} + S_{+}^{\prime} =  \frac{\pi \sin \pi t_0}{2} 
(-1)^{m+1} ( U_{m-K_{\Delta}} + U_{m+K_\Delta}) + O_1 \; , 
\end{equation}
with $O_1 \sim O(N^{-1})$. Note that in the case of an odd N, the residual 
$O_1$ also includes 
 the difference between the last term, $k=N$, and one {\it half} of
the 
$k=(N-1)$-th term, which is of order $U_{N}^{\prime} \sin(\pi t_0) /N$.
Here we take into account that $C_{N,N-1}=2\, C_{N,N} $ from 
\eqref{classsize}.
Thus, for any $\epsilon$ we can chose   $N_\epsilon$ such that
for all $ N > N_\epsilon$ the residual in \eqref{Spmprime}  
is $O_1 < \epsilon/2$. Then \eqref{fprime} is reduced with accuracy 
$< \epsilon / 2$  
to a summation of elements localized around, and 
symmetric with respect to,  the point $t_0= m/N $:
\begin{eqnarray}\label{fprimelocal}
f_{N}^{\prime}(t_0) & = & \pi \sin \pi t_0 \sum_{j=1}^{K_\Delta-1}
(-1)^{j} (U_{m-j} +U_{m+j}) +  
 \nonumber \\
& & \frac{\pi \sin \pi t_0}{2}  (-1)^{K_\Delta} 
 ( U_{m-K_{\Delta}} + U_{m+K_\Delta}) 
+ \frac{\pi \cos \pi t_0}{2 \sin \pi t_0} g_{m}  
 + O_1( \epsilon /2) \; . 
\end{eqnarray}
Here we take into account that $K_\Delta$ is chosen to be of the same
parity with $m$.
Introducing now $r_j = j/N$ which is $\leq \Delta$ for $j\leq
K_\Delta$,
we have
$$ 
\sin \pi t_0 (U_{m-j} +U_{m+j}) = g_m \frac{\cos \pi t_0}{\sin \pi t_0}
 - \frac{2}{\pi} g_{m}^{\prime} + O_{2,j} \; ,
$$
 where, keeping the largest order terms,
the residual $O_{2,j}$ is reduced to
\begin{equation}\label{O2j}
O_{2, j} \simeq \left( g_m - \frac{2}{\pi} g_{m}^{\prime} + 
\frac{2 \sin^{2}\pi t_0}{\pi^2}\, g^{\prime\prime}_{m} \right) 
\left( \frac{\pi r_j}{2 \sin \pi t_0} \right)^2 \; . 
\end{equation} 
Substituting these 2 relations into \eqref{fprimelocal}, and recalling
that 
$g_{m}^\prime \equiv g^{\prime}(t_0=m/N)$, it is easily shown that for both 
 even and odd $m$ and  $K_\Delta$ one has:
\begin{equation}\label{fprimefin}
f_{N}^{\prime}(t_0)= g^{\prime}(t_0) +O_{1}(\epsilon/2) + O_2(\Delta^2)
\; ,
\end{equation}
where the residual $O_2$ represents the sum of  
residuals $O_{2,j}$, i.e $O_2\sum_{\{j \} } O_{2,j}$. 
Because of the sign alteration term $(-1)^{j}$
in \eqref{fprimelocal}, this sum does not increase with increasing 
$N$ beyond the absolute value of $O_{2,j}$ at $r_j =\Delta$
in \eqref{O2j}. It follows then that for any small $\epsilon > 0 $ we can 
first choose an interval $\Delta$ proportional to $\sqrt{\epsilon}
\, \sin \pi t_0$ (depending also on $g, \, g^\prime, \, g^{\prime
\prime}$) such that $\mid O_2 \mid < \epsilon/2$. Then we can choose a number 
$N_\epsilon$ such that $\mid O_1 \mid < \epsilon/2$. Hence 
$\mid O_1 +O_2 \mid < \epsilon$  for all $N > N_\epsilon$. This proves 
the differentiation Theorem.
\hfill$\blacksquare$ 
 
\vspace{3mm}

Note that in the derivation of \eqref{Uestimate} which has allowed
principal
clipping of both end-terms in the localized \eqref{fprimelocal} exactly by
one 
half, the symmetry properties of the \su2 DGT series expressed in the term   
$C_{N,k}$ of \eqref{classsize} have been fully exploited.

Another notice is that, along with the continuous DGT extension of
$\{g_k\}$, the {\it localization Lemma is also valid for its term-by-term
derivative} $f_{N}^{\prime}(t)$. This property is actually 
proven by the localized structure of the sum in 
the construction \eqref{fprimelocal}. 
\smallskip

\vspace{4mm}

\noindent
{\bf Example~4.}
In Fig.~4 we show approximations to the function 
\begin{equation}\label{figure4}
g(t) = e^{-4t}+\frac{1}{2}e^{-\frac{1}{2}\left(
\frac{t-0.5}{\sigma}\right)^2} 
\end{equation}
provided by the continuous extensions of the discrete Fourier 
transforms in forms of the series \eqref{CEDGT} for \su2 DGT and the series
\eqref{approximate}  based on the first $j\leq N/2$ 
half of the standard DFT coefficients $u_j$ of \eqref{stDFT}.
It is noteworthy to recall that a straightforward  use of the 
continuous extension of the standard DFT given by \eqref{CEDFT} for  
calculations of derivatives does not make any sense 
as far as such an extension does not converge, as shown 
in Fig.~2.    
Although the DFT series $s_N(t)$ does converge to $g(t)$ with increasing
N, as shown on the top right-side panel, its derivative series  
contains profound oscillations at any large N. One could suggest
that $g^\prime(t)$ can be in principle recovered from $s_{K}^{\prime}(t)$
after the application of some smoothing procedure. Although this might be 
possible, special care should be excercized in order to avoid
accumulation
of systematic errors in circumstances where the amplitude of the
oscillations is much higher than the mean expected value. Meanwhile,
$f_{N}^{\prime}(t)$ shown by solid lines on the two bottom panels of
Fig.~4 already provides a rather good approximation to $g^{\prime}(t)$ at 
relatively low values of $N$. 

At the end of this section we would just like to note, albeit without
any proof or demonstration in this paper, that our numerical 
calculations show that the second derivative of CEDGT also appears 
to converge to $g^{\prime \prime}(t)$ if
the condition $g(0) = g(T_0)$ is satisfied.

\section{Multidimensional Fourier 
transform }

Being a transform with separable variables, the multidimensional
DCT is easily reduced to the product of one-dimensional DCTs, which is widely
used, for example, for effective 2D-image processing 
(e.g. see \cite{DCT,Strang}.
Although the multidimensional DCT is well known, in this section we will first
formulate it in terms of discrete Fourier transform on the
  $SU(2)\times\cdots\times SU(2)$ group. Then we will briefly 
discuss the convergence properties of the continuous extension of 
2D discrete cosine transforms. 
  
The generalization of the transform formulae 
for decomposition of functions of $n$ variables, $G(x_1,\dots,
x_n)$, into the  Fourier series of orbit functions of $[\su2]^n$
group is straightforward. Let us consider  first the case of $n=2$, 
i.e. when a function  $G \rightarrow G(x,y)$ defined in the region
${\sf F_n} = \{0\leq x\leq 1,\ 0\leq y\leq 1\}$ (i.e.
assuming normalized variables $x\rightarrow x/X_0 $,
$y\rightarrow  y/Y_0 $) is to be decomposed  into the series of the
orbit functions $\Phi_{m n} (x,y)$ of the  symmetry group $\su2
\times \su2 $. In this case  
$\Phi_{m n}(x,y) = \Phi_m(x) \,\Phi_{n}(y)$. Using for convenience
again the functions $\psi_{m}(x) = \cos (\pi m x)$ instead of
$\Phi_m(x)$, we can write
\begin{equation}\label{orbit2}
\Psi_{m n}(x,y) = \psi_{m}(x) \psi_{n}(y) = \cos(\pi m x) \cos(\pi n y) \,
,
\end{equation}
where $(x, y) \in  F $. For a uniform rectangular grid 
$\{x_j, y_k\}$ defined in the region {\sf F} such that
$$\{ x_j = j/M, \; y_k =k/N \; \mid j = 0, 1, \ldots , M \; ; \; 
                   k= 0, 1, \ldots , N \} $$      
the functions $\Psi_{m n}$ are orthogonal in the 
following bilinear form   
\begin{eqnarray}\label{orthogon2}
\langle \Psi_{m n} , \Psi_{p q} \rangle_{M, N} & = & 
\sum_{j=0}^{M} \sum_{k=0}^{N} C_{M,j} C_{N,k} \Psi_{m,n}(x_j,y_k)
\Psi_{p q}(x_j,y_k) \nonumber \\
 & = &\frac{ 4 M N  }{ C_{M,j} \, C_{N, k} }\delta_{m p} \, \delta_{n q}
\; ,
\end{eqnarray}
which follows directly from \eqref{orthogonality}.

Let $G(x,y)\rightarrow\R$ be a continuous function originating
a 2-dimensional grid function  $G_{jk} = G(x_j,y_k)$ on the  
grid $\{x_j,y_k \}$. Then the decomposition of $G$ into the Fourier series on 
$\su2\times \su2 $ group corresponds to solving the system of equations
\begin{equation}\label{Gjk}
  G_{jk}  =  \sum_{m=0}^{M}\, \sum_{n=0}^{N} A_{mn}\,
\Psi_{mn}(x_j, y_k) 
\equiv \sum_{m=0}^{M}\,\sum_{n=0}^{N} A_{mn} \,
 \cos \frac{\pi mj}{M} \, \cos \frac{\pi nk}{N}\; ,   
\end{equation}
with $\{ 0\leq j \leq M \, , \, 0 \leq k \leq N \}$, with respect
to the coefficients $A_{mn}$. This can be easily achieved using the
orthogonality relation \eqref{orthogon2}, if we multiply  
\eqref{Gjk}  by $C_{M,j} C_{N, k} \Psi_{p q} (x_{j},y_{k})$ and 
take the sum over $\{j,k\}$. Thus we find the
coefficients $A_{m n}$ of the discrete Fourier transform \eqref{Gjk},
\begin{equation}
A_{mn} = \sum_{j=0}^{M}\sum_{k=0}^{N}D_{M}^{mj}D_{N}^{nk}G_{jk}\,.
\end{equation}   
The matrix ${D}_N$ is defined as before by \eqref{Djk}, 
$$
D_{N}^{a b} =\frac{C_{N,a}C_{N,b}}{2N}\cos\frac{\pi ab}{N}\,,
\qquad N,a,b\in\mathbb Z\,,
$$ 
with the weights  $C_{N,a}$, $C_{N,b}$ given by \eqref{classsize}. 

In this way the coefficients $A_{mn}$ of the 2-dimensional DGT are
found for any grid function $\{ G_{jk} \}$ with bounded
values at the grid points
$(x_j, y_k)\in\sf F$. Thus we can formulate the following proposition:
\medskip

\noindent 
{\bf Proposition:} {\it Let $G_{jk}= G(x_j,y_k)$ be values of 
a bounded function $G(x,y)$ on the rectangular grid points  
$$
x_j=jX_{0}/M,\quad y_{k}=kY_{0}/N\,;\qquad 
j\in\{0,1,\ldots,M\}\,,\quad k\in\{0,1,\ldots,N\}\,.
$$
A trigonometric function given by finite Fourier series
\begin{equation}\label{CEDGTtwo}
F_{M N}(x,y) = \sum_{m=0}^{M} \, \sum_{n=0}^{N}  \, A_{mn} 
\cos \frac{\pi m x}{X_0} \, \cos \frac{ \pi n y }{ Y_0 } \; ,    
\end{equation}
with 
\begin{equation}\label{DGT2}
A_{mn}=\sum_{j=0}^{M}\sum_{k=0}^{N} 
\frac{C_{M,m}C_{M,j}C_{N,n}C_{N,k}}{4MN}
G_{jk}\cos\frac{\pi mj}{M}\cos\frac {\pi nk}{N}\;,
\end{equation}
continuously extends the discrete inverse Fourier transform 
of the grid function $\{ G_{jk}\}$ onto the entire rectangular area
$(x\in [0,X_0], y\in [0,Y_0] )$, and  
satisfies the equality
$$
F_{M N}(x_{j},y_{k})=G_{jk}
\quad\text{for all}\quad 
j\in\{0,1,\ldots,M\}\,,\quad k\in\{0,1,\ldots,N\}\,.
$$

\noindent
Furthermore, if $G(x,y)$ is continuous, then $F_{MN}(x,y)$ 
converges to $G(x,y)$ for $M,N\longrightarrow\infty$.
}
\medskip

Since for any fixed $y_0 \in [0, Y_0]$ or $x_0 \in [0, X_0]$ the 
series $F_{MN}(x,y_0)$  or $F_{MN}(x_0, y)$, respectively, are reduced to 
one-dimensional CEDCT/CEDGT series along the $x$ or $y$ axes considered in 
the previous section, it is obvious 
that the continuous extension $F_{MN}(x,y)$
of 2D DGT on the SU(2)$\times$SU(2) group (i.e. the 2-dimensional DCT)
 not only converges with increasing $(M, N)$ to $G(x,y)$,
but also that it has properties of {\it locality} and
{\it differentiability} similar to the one-dimensional 
CEDGT on ordinary \su2\, group.

\medskip

\noindent
{\bf Example~5.} The upper panels in Fig.~5 show the contour plots 
of a  function
$G(x,y)$ defined in the square region ${ F} = [0,1]\times[0,1]$
and composed  of two 2-dimensional Gaussian functions,
each of type 
\begin{equation}\label{figure5}
   e^{-\frac{(x^\prime-x_0)^2}{2 \sigma_{\parallel}^2 } -
      \frac{(y^\prime-y_0)^2}{2 \sigma_{\perp}^2 } } \; , 
\end{equation}
where $\sigma_{\parallel} \geq \sigma_{\perp}$, but with 
directions of the major axes $x_{1}^\prime$ and
$x_{2}^\prime$ perpendicular to each other.  For both Gaussians we
have taken the transverse dispersions to be  
$\sigma_{1,\perp}=\sigma_{2,\perp} = 0.025$, which is exactly 2
times smaller than the grid's cell size $\Delta x = \Delta y = 1/20$
for the chosen $M=N=20$.  The contour plots shown on the upper 
panels in Fig.~5 illustrate that even in the case of a
grid  with cell size this large compared with $\sigma_{\perp}$, the
continuous extension of the two-dimensional DGT series reconstructs  
Gaussian-fast smooth structures reasonably well. 

This is also apparent on the bottom panels of Fig.~5 where 
we show the 3-dimensional images for the same 
analytic function $G(x,y)$ (left panel) and its approximation in the form of
2-dimensional continuous DGT extension $F_{MN}(x,y)$ (right panel). 
Note that any waviness 
that can be seen in the approximated function 
would disappear from the images had we taken $N,\, M
\geq 1/\sigma_{\perp}$ for the same functions.

\smallskip

\noindent
{\bf Example~6.}~~The latter case is chosen in Fig.~6 where 
the upper panels show, in terms of brightness distribution,
the original grid function produced by 2 Gaussians with 
$\sigma_{1,\perp}=\sigma_{2,\perp} =0.05 =1/N$, for $N=M=20$,
 and its reconstruction in the form of 2-dimensional continuous 
DGT extension (on the right). The major axes of the ellipsoids
are inclined at a small angle ($20^\circ$) to each other. 
For comparison, on the bottom left panel we show the contour plot
of the exact (i.e. originating) function $G(x,y)$, and the bottom
right panel shows its approximation by 2-dimensional CEDGT.
It is obvious that  the reconstructed CEDGT image not only 
recovers the directions of the ellipsoids and their maxima,
but it practically coincides
with the exact image. Note that the dashed contours on both Fig.~5 and
Fig.~6 show a level slightly below zero, $F_{MN}= -0.001$.

\medskip
Generalization of the proposition for an n-dimensional DGT of a 
function $G(x_1, \ldots , x_n)$  on $ [ {\bf SU} (2) ]^n$ group is
straightforward:
\medskip

\noindent     
{\bf Proposition:}      
  {\it   Let $G_{j_1 \ldots j_n} = G(x_{1, j_1}, \ldots , x_{n, j_n}$ 
be values of a bounded function $G(x^{(1)}, \ldots , x^{(n)})$ given on 
the rectangular $(M_1 , \ldots , M_n)$-interval grid points 
$$\{ x^{1,j_1} = j_{1} X_{1}/M_1 , 
\ldots , x^{n,j_n}=j_n X_{n}/M_n \, , \qquad   
j_k  = 0, 1, \ldots , M_k \, : \; k = 0, 1, \ldots , n   \}  \; .$$ 
A continuous function given by finite Fourier series
 \begin{equation}
F_{M_1 \ldots  M_n}(x_1, \ldots , x_n) = 
\sum_{m_1=0}^{M_1}  \ldots  \sum_{m_n=0}^{M_n}  A_{m_1 \ldots m_n} 
\cos \frac{\pi m_1 x_1}{X_{1}} \cdot \ldots \cdot  
\cos \frac{ \pi m_n x_n }{ X_{n} } \; ,    
\end{equation}
where 
\begin{eqnarray}
A_{m_1 \ldots m_n} & = & \sum_{j_1 = 0}^{M_1} \ldots \sum_{j_n=0}^{M_n} 
\frac{C_{M_1,m_1} C_{M_1,j_1} \cdot \ldots 
\cdot C_{M_n,m_n} C_{M_n,j_n} }{2^{n} 
\, M_1 \cdot \ldots \cdot  M_n} \times \nonumber \\  
 &  &  G_{j_1 \ldots j_n} \, \cos \frac{\pi m_1 j_1}{M_1} 
\cdot  \ldots \cdot   
 \cos \frac{\pi m_n j_n}{ M_n}  
\end{eqnarray}
satisfies the equality
$$F_{M_1 \ldots M_n}(x_{j_1}, \ldots , x_{j_n}) = 
G_{j_1 \ldots j_n} \; , \qquad {\rm for \; all} \; 
 j_k \in \{0, 1, \ldots , M_k\} \; {\rm and} \;  
 k \in \{ 0, 1, \ldots , n \} \; .$$
\vspace{2mm}

\noindent
Furthermore, if $G(x_1,\ldots , x_n)$ is continuous, then 
$F_{M_1 \ldots M_n}(x_1, \ldots , x_n)$ 
converges to $G(x_1,\ldots x_n)$ with 
$M_1, \ldots , M_n \longrightarrow \infty$.  
} 

\vspace{3mm}
The proof of this proposition is readily obtained by the method of induction
on the SU(2) factors of the group.

\smallskip

\subsection{An example of CEDCT application to real images}

In order to demonstrate the potential of the 
above suggested approach 
of {\it continuous extension} of the inverse multi-dimensional discrete
group transforms for purposes of natural interpolation of discrete images
between the grid points, as well as for 
the possibility of data compression and smooth representation of   
the compressed images, we have chosen a 56$ \times $140 pixel
fragment of the well known image ``Lena''. The original fragment
shown on Fig.~7a is strongly enlarged (`zoomed')
in order to make visible the granularity of the image at 
its resolution limits. The grayscale color coding of the fragment contains  
all 256 intensity levels, from $g=0$ ({\it black}) to $g=255$ 
({\it white})

In Fig.~7b we show the continuous extension 
of the original image. It is constructed by subdividing each of the initial
intervals $\Delta x$ and $\Delta y$ into 3 subintervals. 
This procedure increases the density of the grid points (pixels) 
by a factor $3\times3=9$. Calculations are done dividing first  
the initial 56$\times $140 pixel fragment into ten 28$\times $28 pixel
subfragments, then calculating the CEDCT for each of these sub-fragments.
It allows us to demonstrate, on the 
next two panels,  the effects of CEDCT image reconstruction
at the block edges after some image compression is done. 
Because the continuous extension of the inverse DCT
can formally extend the values of the initial intensity distribution function
to values $F_{MN}(x,y)$ beyond the limits $[0,255]$ used for the 
intensity coding, we have  linearly renormalized  the values of 
$F_{NN}\rightarrow \tilde{F}_{NN}\in [0,255]$. The positive impact 
of the higher resolution achieved by the use of CEDCT in Fig.~7b, 
as compared with the original image in Fig.~7a, is apparent.

Note that the harmonic order of the cosine functions 
$\cos (\pi n x/X_0$ and  $\cos (\pi m y/Y_0$ used in CEDCT corresponds to
the modes $0\leq m,n\leq 50$. In Fig.~7c we show the continuous 
extension of the image obtained after application of the simplest  
``low-pass compression'' procedure (see e.g. \cite{DCT}), putting the DCT
coefficients $A_{mn}\rightarrow 0$ for all high-order modes with either $m$ or
$n$ exceeding $n_{max}=19$. This procedure generally removes the high-frequency
`noise' from the image, and compresses the image by a factor
$(29/20)^2\approx 2$. No visual degradation of the CEDCT image is apparent. 
It is noteworthy that although the exactness property of the transform
in Fig.~7c is lost, the edges of the 10 individual blocks, or the 
sub-fragments, cannot be visually distinguished. 
The block edges become noticeable only in 
 Fig.~7d, where we have applied again the CEDCT approach for visualization of 
the image compressed now by a factor of 10. The compression in Fig.~7d is
effectively reached by keeping in CEDCT series only $10\%$ of the cosine 
terms with the large-value coefficients $A_{mn}$, and discarding all 
terms with small-value $A_{mn}$. For Fig.~7b it has corresponded to the 
assumption $A_{mn} \rightarrow 0 $ if $\mid A_{mn}\mid \leq 0.05\,
 A_{\rm max}$, where $A_{\rm max}$ is the maximum absolute value of the
coefficients $\{ A_{mn}\mid \}$ ({\it excluding} $A_{00}$). 
Obviously, the image in Fig.~7d still remains smooth and quite 
recognizable.

We would like to note here that it has not been our aim in this paper to reach 
the goal of the best possible 
image compression. We believe, however, that through this paper
our examples
demonstrate the high potential of the developed approach of 
{\it continuous extensions} of DCT, and of the DGTs generally, as 
considered in the next Paper II for the SU(3) group, for purposes of 
practical applications, and in particular for image processing and 
compression.

\section{Summary}

We have shown that:

\medskip 

\noindent 
1.~~~A discrete Fourier transform of a grid function $\{ g_k \mid 
k=0, 1, \ldots , N\}$ on the orbit functions of Lie groups,
abbreviated DGT, in the case of \su2 is reduced to the well known 
discrete cosine transform, namely to DCT-I, 
which is a known type of exact discrete transforms,
like the standard DFT sequence.  
\vspace{2mm}

\noindent
2.~~~The principal difference
between these 2 types of discrete Fourier transforms consists in the fact 
that DCT is based on the functions $\cos (k \pi t/T_0)$ 
corresponding to $k\leq N$ trigonometric harmonics of both integer 
and half-integer orders, $n=k/2 \leq N/2$, whereas the DFT   
utilizes trigonometric functions of integer $n$ only, but
extending to orders $n\leq N$. This results in vital differences 
in the subsequent properties of DFT and DCT (or DGT generally).       
\vspace{2mm}

\noindent
3.~~~If the function $g(t)$ originating $\{g_k\}$ is a continuous function 
of $t\in [0,T_0]$, then the continuous extension
of the  ({\it inverse}) DCT sequence results in the function $f_N(t)$ 
which converges to the original $g(t)$ with increasing $N$ at all 
$t$. This property does not hold for the  
continuous extension of the standard DFT sequence, which shows  
profound oscillations between the points of the grid. Note
that potentially 
this feature implies significantly smaller vulnerability of DCT to the
truncation/approximation errors in the process of filtering 
as compared with the standard DFT. Therefore it could be the reason 
for the superior general performance of the DCT compared with the DFT 
(see \cite{Rao1}).  
\vspace{2mm}

\noindent
4.~~~Similar to canonical {\it continuous} Fourier transform 
polynomials with coefficients calculated by exact integrations,
the CEDCT series satisfies the principle of {\it locality}. 
This property insures, in particular, 
that the computation errors connected, e.g.,
with noise or uncertainties in one segment of the data will not 
significantly affect 
the reconstructed CEDCT image on the distant segment of data.
This property of DCT may become important especially in 
the process of lossy data compression when the property of exactness
of the {\it discrete} transform is not necessarily preserved.

\vspace{2mm}

\noindent
5.~~~Similar to the canonical CFT, the CEDCT series $f_N(t)$ 
can be differentiated term by term, so that for the ({\it first})
derivative
series $\lim_{N\rightarrow \infty} f_{N}^{\prime}(t) \rightarrow 
g^{\prime}(t)$ for all $t\in (0,T_0)$ provided that the second
derivative of $g(t)$ is  
a continuous (or just an integrable) function on the interval
$[0,T_0]$.  For CEDCT this property is valid both when $g(0)=g(T_0)$
 and $g(0)\neq g(T_0)$.
 It does not necessarily hold for 
other types of discrete Fourier transforms which might themselves be 
converging, like $s_k(t)$ in \eqref{approximate},  
but which produce nontheless a non-converging derivative series, 
as demonstrated in Fig.~4.    

The derivative series $f_{N}^{\prime}(t)$ satisfies the localization
principle along with $f_N(t)$.

\vspace{2mm}

\noindent
6.~~~In the case of an $n$-dimensional function defined on the 
knots of a rectangular $n$-dimensional grid, the DGT
Fourier decomposition can be performed using
the orbit functions of $[SU(2)]^{n}$ group. Such Fourier 
series are effectively reduced to the $n$-fold convolution
of one-dimensional DGT on SU(2) alongside $n$ independent
(rectangular) axes, and therefore they posess nice 
properties of {\it convergence}, {\it localization}, and 
{\it differentiability}
of their continuous DGT extensions similar to the one-dimensional
CEDCT.   
 
\bigskip

\section*{Acknowledgement}

The authors acknowledge partial support of the National 
Science and Engineering Research Council of Canada, of FCAR of
Quebec, and of NATO.



\begin{figure}
\centerline{\epsfig{file=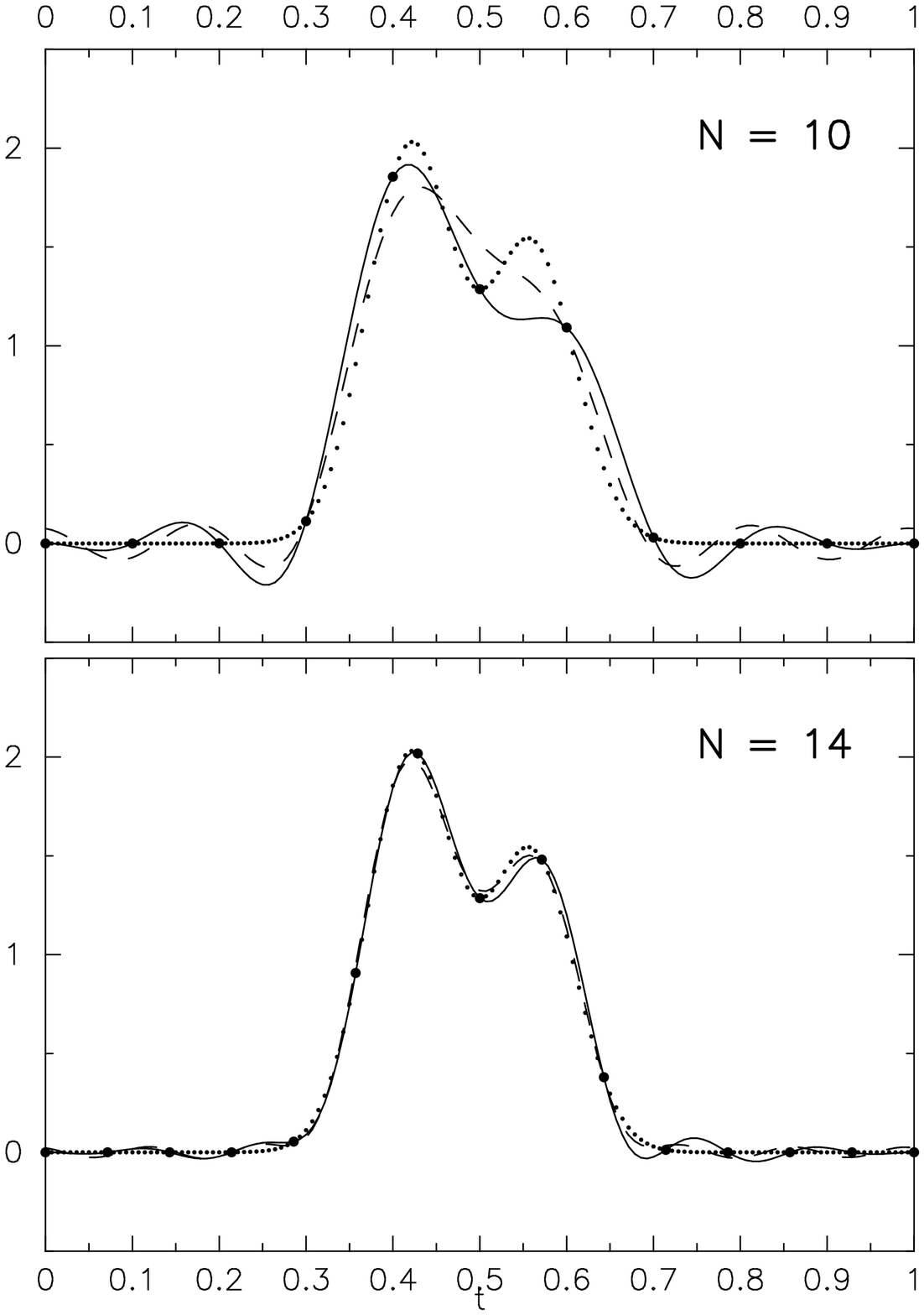, width=7.7cm}}
\caption{Approximation of an analytic function 
\eqref{figure1}, shown by the  
dotted lines, with the CEDCT function $f_N(t)$ (solid lines) for the discrete 
interval numbers 
$N=10$ and $N=14$. The big dots show the values of the grid function
$\{ g_{k}\equiv g(t_k)\}$ for $\{ k= 0, \ldots , N \}$. For comparison we
also show
by dashed lines the approximation of $g(t)$ by the exact 
CFT polynomials of \eqref{exact}
with $K=N/2$. It gives polynomials of the same harmonic maximum order as 
the ones in the DCT. The parameters in \eqref{figure1} are:  $A1=2$, 
$A_2=1.5$, $\sigma_1=\sigma_2= 0.05$,  $t_1=0.42$, $t_2 =0.56$.}
\end{figure}

\begin{figure}
\centerline{\epsfig{file=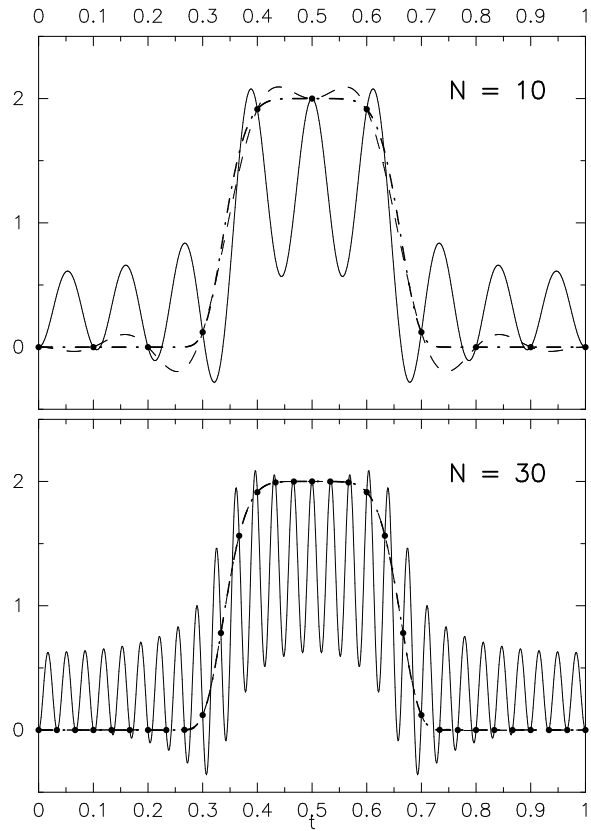, width=7.7cm}}
\caption{Behaviours of the Fourier series $h_N(t)$ and $f_N(t)$,
given by equations \eqref{CEDFT} and \eqref{CEDGT}. Solid lines 
represent the CEDFT, and dashed lines show the CEDCT. 
Big dots show the values of the grid function $\{ g_k \mid k = 0, 
\ldots , N\}$ originated from an analytic 
function $g(t)$ which is given by equation \eqref{figure2},
and is shown by the dot-dashed line.}
\end{figure}

\begin{figure}
\centerline{\epsfig{file=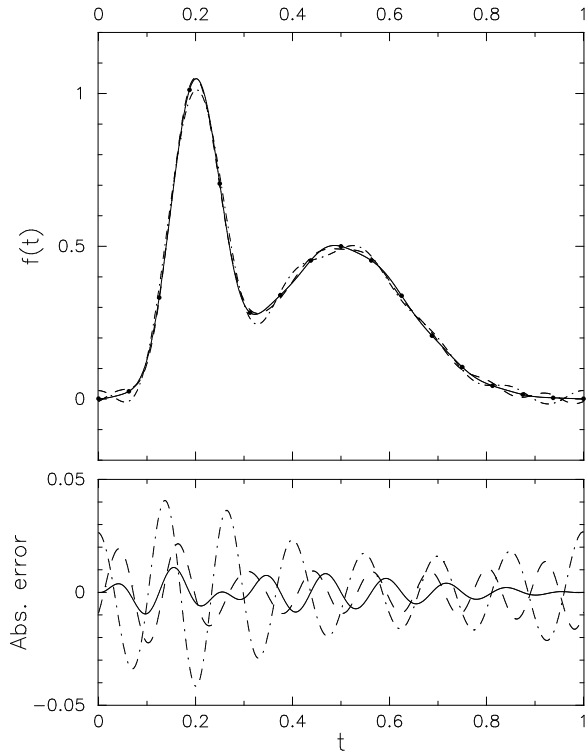, width=7.7cm}}
 \caption{Approximation of an analytic function $g(t)$ composed of two 
Gaussians, \eqref{figure1}, with the CEDCT function 
$f_{N}(t)$ of \eqref{CEDGT}
with $N=16$ ({\it solid lines}) and the Fourier series $s_{K}(t)$ of 
\eqref{approximate} that uses 
only the first $K\sim N/2$ coefficients $\{ u_{j} \mid j \leq K \}$  
of the standard DFT \eqref{stDFT}. 
The dashed line is for $K=N/2 = 8$, and the dot-dashed 
line corresponds to $K=6$. The full dots in the upper panel correspond to   
$\{ g_{k} \mid k = 0, \ldots , N \}$. The lower panel shows the
corresponding 
errors of the approximation to $g(t)$ by these 3 types of 
discrete Fourier transforms. The 
dispersions in \eqref{figure1} are assumed to be $\sigma_1= 0.07$ and 
$\sigma_2 = 0.2\,$.}
\end{figure}

\begin{figure}
\centerline{\epsfig{file=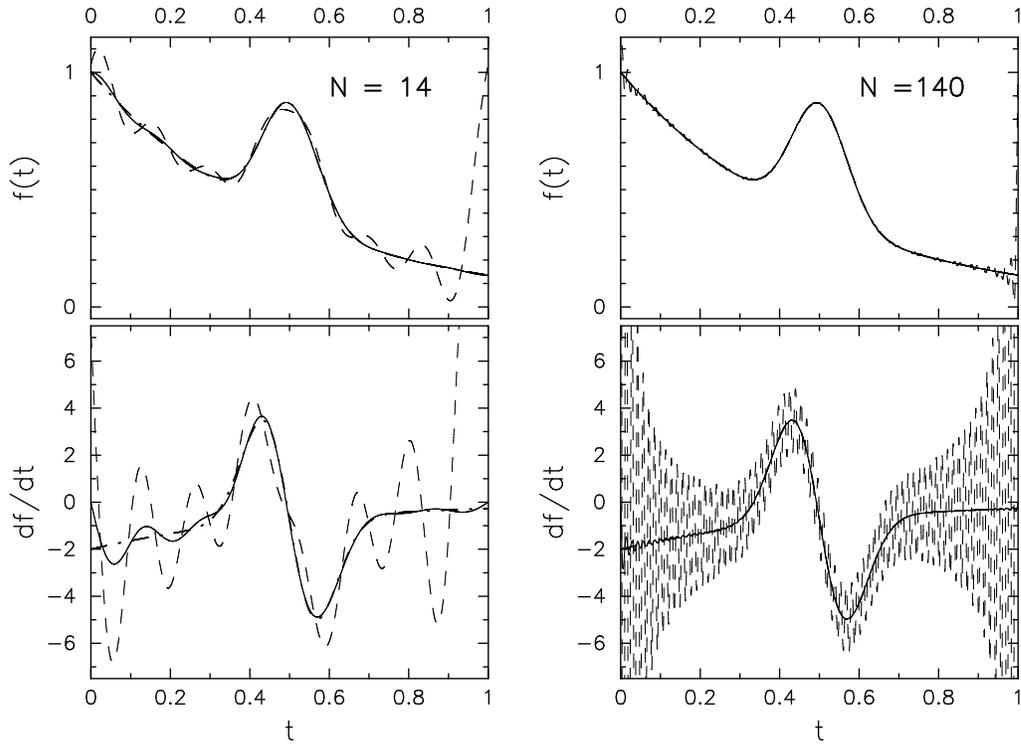, width=13.5cm} \hspace{2mm} }
\caption{The CEDCT series $f_{N}(t)$ ({\it solid lines, top panels}) and the
derivatives $f_{N}^{\prime}(t)$ ({\it solid lines, bottom panels}) for  
an analytic function $g(t)$ of equation \eqref{figure4} and its derivative
$g^\prime(t)$ ({\it heavy dot-dashed lines}) in the case of $N=14$ and
$N=140$.
For comparison, by dashed lines we show the truncated DFT series 
$s_{K}(t)$ of equation \eqref{approximate} and its derivative 
$ s_{K}^{\prime}(t)$ with $K=N/2$. Note that although $s_{K}(t)$  
converges with $N\rightarrow \infty$ at all $t\in (0,1)$,  its 
derivative $s_{K}^\prime$ apparently does not.}
\end{figure}

\begin{figure}
\centerline{\epsfig{file=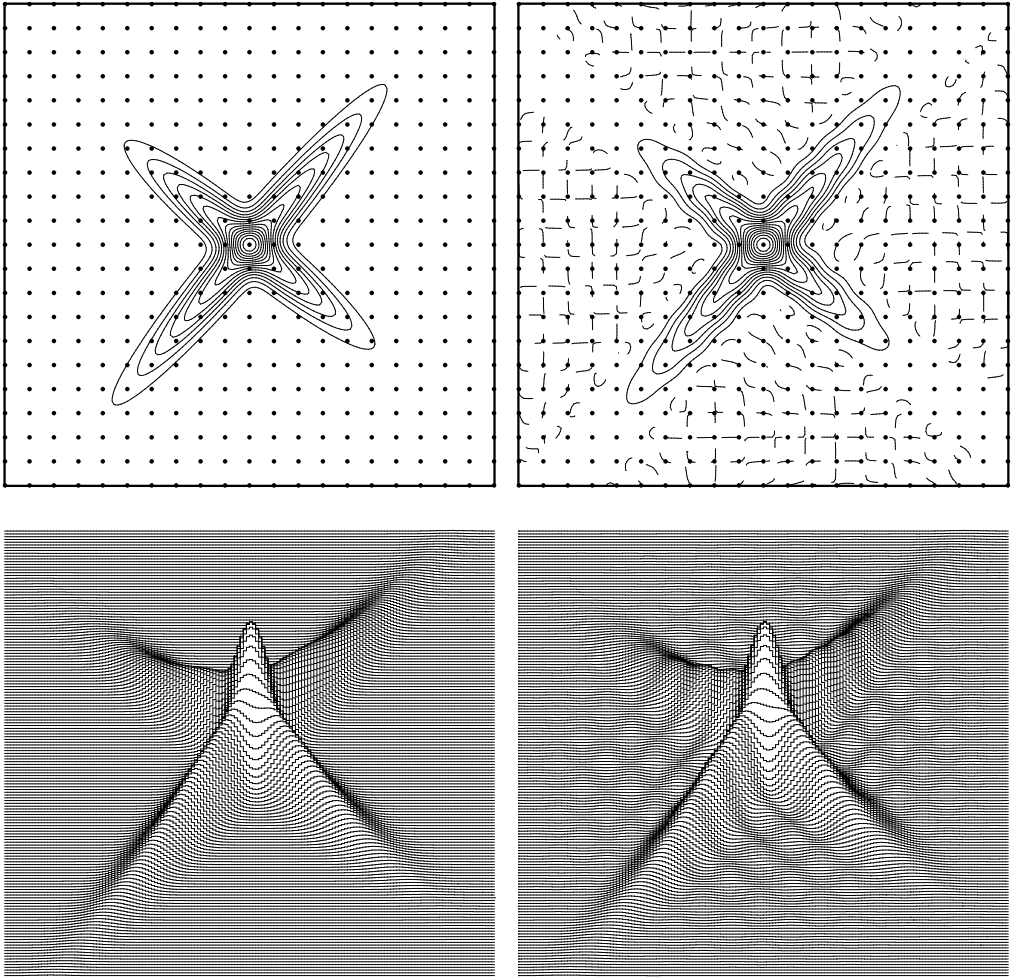, width=13.5cm} }
\caption{
 Contour plots (upper panels) and visual `3D' images (lower
 panels)
 of an analytic function $G(x,y)$ (right side) and its approximation by
 2-dimensional  \su2$\times$\su2  CEDGT/DCT series $F_{MN}(x,y)$ 
 (left side) with $M=N=20$.
 $G(x,y)$ is composed of a sum of 2 two-dimensional
 Gaussian ellipsoids, each in the form of \eqref{figure5}, and with  
 dispersions $\sigma_{\perp, 1} = \sigma_{\perp, 2} = 0.025=1/2N$. 
 The dashed lines on the left bottom panel show the contour level 
 $F_{MN} (x,y) = - 0.001$\,.
}
\end{figure}

\begin{figure}
\centerline{\epsfig{file=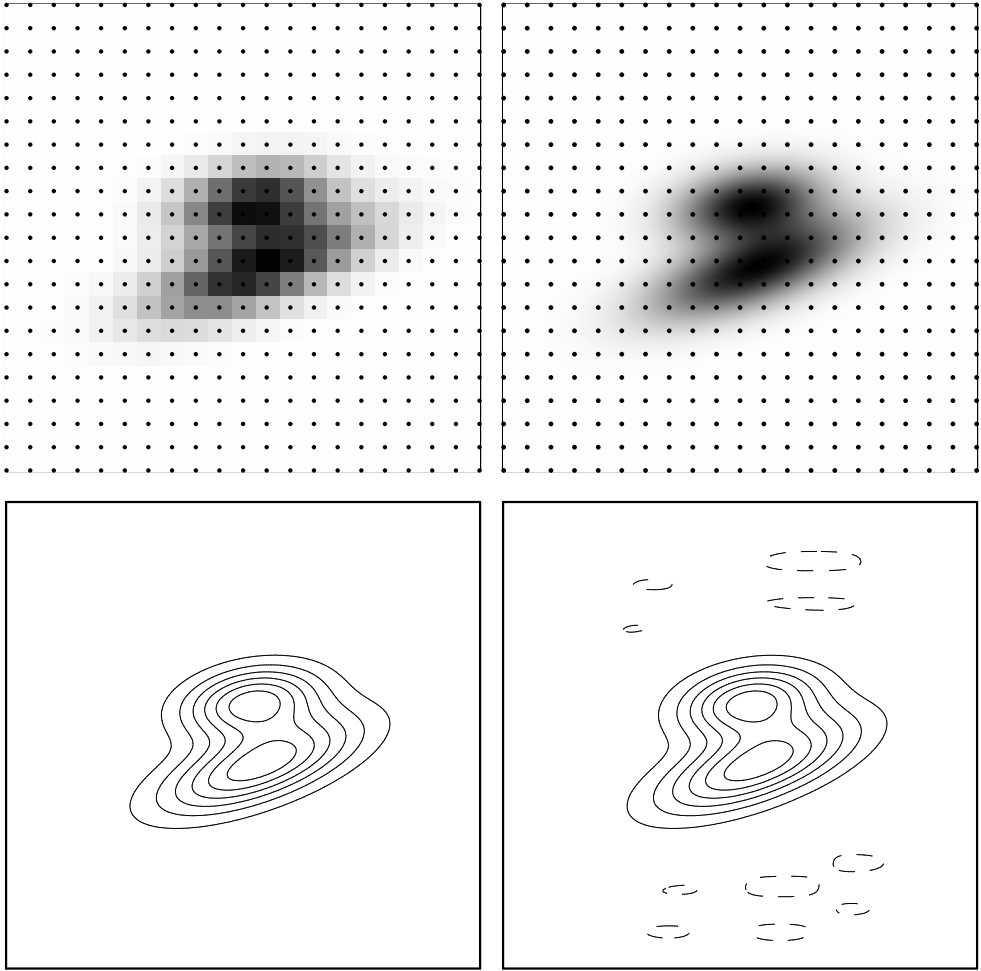, width=13.cm} }
\caption{ {\protect \bf Upper panels:} The images, in terms of
brightness distributions, corresponding to the grid function 
$\{G_{j,k}\}$ given on $M=N=20$ square grid (on the left), and its 
reconstruction by 2-dimensional CEDCT series $F_{M,N}(x,y)$ (on the right).   
{\protect \bf Lower panels:}  Contour plots of the analytic
function $G(x,y)$ originating $\{ G_{j,k}\}$ 
(on the left), and of its approximation by $F_{M,N}(x,y)$ (on the right).
The original function $G(x,y)$ is composed of two-dimensional Gaussian 
distributions with transverse dispersions for both $\sigma_{\perp}=1/N=
0.05$, an angle between long axes of the ellipsoids
equal to $20^\circ$, and a small separation between  their peak positions.}
\end{figure}

\clearpage

\begin{figure}
 \centerline{\epsfig{file=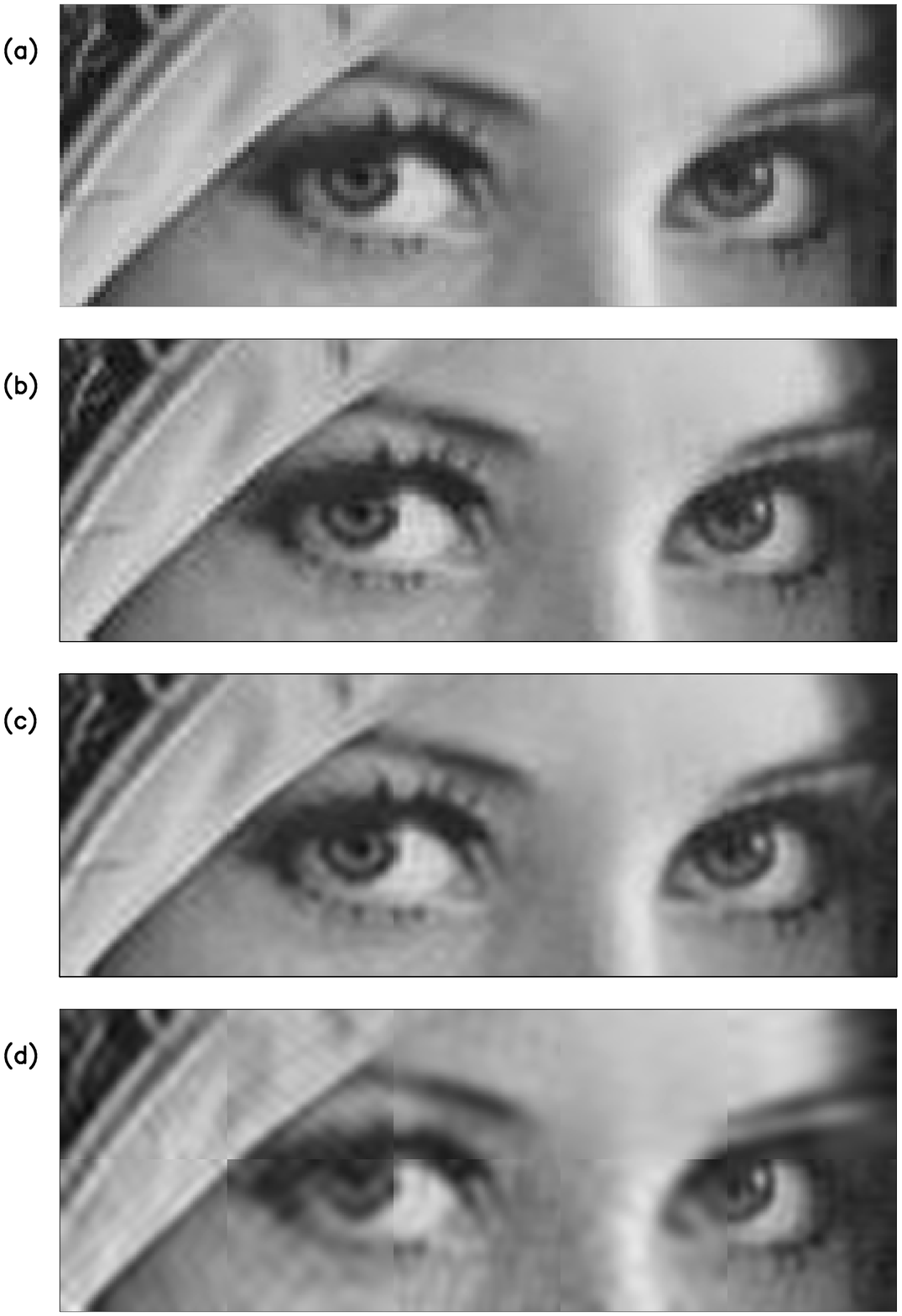, height=20.5cm} }
\end{figure}

\begin{figure}
\caption{ {\protect \bf (a)} Zoomed $56\times 140$-pixel 
fragment of the original image ``Lena'';
          {\protect \bf (b)} CEDCT image calculated with the
3$\times$3 higher resolution, i.e. corresponding to $168\times 420$ pixels;
For our calculations the image was first subdivided into 10 square blocks
of $28\times 28$ size. 
{\protect \bf (c)} The picture in the same  $168\times 420$ pixel 
representation compressed by a factor of 2. The compression 
is reached by retaining   in \eqref{CEDGTtwo}
only the first $0 \leq m,n \leq 19$ DCT coefficients $A_{mn}$, 
and discarding all higher-order terms.
{\protect \bf (d)} The picture compressed by a factor of 10; the compression 
is achieved by discarding all cosine terms with 
$\mid A_{mn}\mid \leq 0.05 \, A_{\rm max} $ (see text).  }
\end{figure}


\begin{thebibliography}{9}

%
\bibitem{Z}
A.~Zygmund, {\it Trigonometric Series,\/} Cambridge University
Press (1959)
%
\bibitem{Tolst}
G. P. Tolstov, {\it Fourier series}, Dover, N-Y (1976) 
%
\bibitem{MP1}
R.~V.~Moody, J.~Patera, {\it  Computation of character
decompositions of class  functions on compact semisimple Lie
groups,\/} Mathematics of Computation {\bf 48} (1987), 799-827
%
%
\bibitem{MP4}
R.~V.~Moody, J.~Patera
{\it 
Elements of finite order in Lie groups and their applications,\/}
XIII Int. Colloq. on Group Theoretical Methods in Physics,
ed. W.~Zachary, World Scientific Publishers, Singapore (1984),
308--318.

%
\bibitem{Rao1}
N.~Ahmed, T.~Natarajan, K.~R.~Rao, {\it Discrete cosine transform},
IEEE Trans. Comput. {\bf C-23} (1974), 90-93
%
\bibitem{Wang}
Z. Wang, {\it Fast algorithms for the discrete W transform and for the discrete 
Fourier transform}, IEEE Trans. Acoust., Speech and Signal Process.
{\bf ASSP-32} (1984), 803-816
%
\bibitem{Strang}
G. Strang, {\it The discrete cosine transform}, SIAM Review {\bf 41} (1999),
135-147 
%
\bibitem{Brigham}
E. O. Brigham, {\it The fast Fourier transform}, Prentice Hall, Englwood 
Cliffs, N.J. (1974)
%
\bibitem{OppenSch}
A. V. Oppenheim, R. W. Schafer, {\it Digital signal processing},
Prentice-Hall, Englwood Cliffs (1975)
%
\bibitem{Nuss}
H. J. Nussbaumer, {\it Fast Fourier transform and convolution algorithms},
Springer-Verlag, Berlin Heidelberg N-Y (1982) 
%
%
\bibitem{DCT}
K. R. Rao, P. Yip, {\it Discrete cosine transform - Algorithms, Advantages, 
Appliucations}, Academic Press (1990) 
%
\bibitem{Kac}
V. G. Kac, {\it Automorphisms of finite order of semisimple Lie Algebras},
J. Funct. Anal. Appl. {\bf 3} (1969), 252-254
%
\bibitem{MP2}
R. V. Moody, J. Patera, {\it Characters of elements of finite order in 
simple Lie groups}, SIAM J. on Algebraic and Discrete Methods {\bf 5}
(1984), 359-383 
%
\bibitem{MMP1}
W.~G.~McKay, R.~V.~Moody, J.~Patera,
{\it Decomposition of tensor products of $E_8$ representations,\/}
Algebras, Groups and Geometries
{\bf 3} (1986), 286-328

\bibitem{MMP2}
W.~G.~McKay, R.~V.~Moody, J.~Patera,
{\it Tables of $E_8$ characters and decomposition of plethysms,\/}
in {\sl Lie algebras and related topics,}
Amer. Math. Society, Providence R.I., 
eds. D.~J.~Britten, F.~W.~Lemire, R.~V.~Moody (1985),
227--264
%
\bibitem{GP}
S. Grimm and J. Patera
{\it Decomposition of tensor products of the
fundamental representations of $E_8$,\/}
in {\sl Advances in Mathematical Sciences -- CRM's 25 Years,\/} 
ed.~L.~Vinet, CRM Proc. Lecture Notes, Amer. Math. Soc.,
Providence, RI, {\bf 11} (1997) 329--355
%
\bibitem{B}
R.~N.~Bracewell, {\it  Numerical transforms,\/}
Science (1990) 697--704
%
\bibitem{CoolTuk}
J. W. Cooley, and J. W. Tukey, {\it An algorithm for the machine
calculation
of complex Fourier series}, Mathematics of Computation {\bf 19} (1965),
 297-301 
%
\bibitem{Gentleman} 
W. M. Gentleman and G. Sande, {\it Fast Fourier transform for fun and
profit},
AFIPS proc. {\bf 29} (1966), 563-578
%
\bibitem{Singl}
R. C. Singleton, {\it A method for Computing the fast Fourier transform
with auxiliary memory and limited high-speed storage},
IEEE Trans. Audio Electroacoust. AU{\bf -15} (1967), 91-97
%
\bibitem{RadBren}
C. M. Rader and N. M. Brenner, {\it A new principle for fast Fourier 
transformation}, IEEE Trans. ASSP-{\bf 24} (1976)   
%
\bibitem{Winograd}
S. Winograd, {\it On computing the discrete Fourier transform},
Mathematics of
Computation {\bf 32} (1978), 175-199
%
\bibitem{NarPet}
M. J. Narasimha, A. M. Peterson, {\it On the computation of the discrete 
cosine transform}, IEEE Trans. Commun. {\bf COM-26} (1978), 934-946 
\bibitem{TsengMil}
B. D. Tseng, W. C. Miller, {\it On computing the discrete cosine transform},
IEEE Trans. Comput. {\bf C-27} (1978), 966-968
\bibitem{Mak}
J. Makhoul, {\it A fast cosine transform in one and two dimensions}, IEEE Trans.
Acoust. Speech, and Signal Process. {\bf ASSP-28} (1980), 27-34  
\bibitem{VetNus}
M. Vetterli, H. Nussbaumer, {\it Simple FFT and DCT algorithms with reduced 
number of operations}, Signal Processing {\bf 6} (1984), 267-278

\end{thebibliography}
\end{document}